\documentclass[twocolumn]{article}

\usepackage[english]{babel}
\usepackage{authblk}

\usepackage[a4paper,top=2cm,bottom=2cm,left=1.5cm,right=1.5cm,marginparwidth=1.75cm]{geometry}

\usepackage[normalem]{ulem} 
\usepackage{csquotes}
\usepackage{amsmath}
\usepackage{amssymb}
\usepackage{braket}
\usepackage{mathtools}
\usepackage{bm}
\usepackage{multirow}
\usepackage{tabularx}

\usepackage{graphicx}
\usepackage{rotating}
\usepackage{quantikz}
\usepackage{adjustbox}
\usepackage{float}
\usepackage{enumitem}

\usepackage[colorlinks=true, allcolors=blue]{hyperref}
\usepackage{url}
\usepackage{orcidlink}
\usepackage[version=4]{mhchem}

\usepackage{calc}
\usepackage{subcaption}

\usepackage[style=numeric-comp, sorting=none]{biblatex}

\renewbibmacro*{issue+date}{%
    \iffieldundef{issue}{}{%
        \printtext[parens]{\printfield{issue}}%
        \setunit*{\addspace}%
    }%
    \printtext[parens]{\printdate}
}
\newcommand{\brm}[1]{\bm{\mathrm{#1}}}

\title{Many-body post-processing of density functional calculations using the variational quantum eigensolver for Bader charge analysis}

\author{
Erik Schultheis$^*$\orcidlink{0009-0007-4728-7124},
Alexander Rehn\orcidlink{0000-0002-9058-4569}, 
Gabriel Breuil\orcidlink{0000-0001-9753-9384}}

\affil{German Aerospace Center (DLR), Institute for Frontier Materials on Earth and in Space, 51147 Cologne, Germany}

\begin{document}
\twocolumn[
    \maketitle
    \vspace{-0.7cm}
    \begin{center}
            \begin{minipage}{0.95\textwidth}
            Quantum chemistry and condensed matter physics are among the most promising applications of quantum computers. Additionally, estimating properties of a material is crucial to evaluate its industrial applications. To investigate charge distributions of weakly and strongly correlated systems, we calculate Bader charges for various periodic systems by solving many-body Hamiltonians using the variational quantum eigensolver and classical solvers. The Hamiltonians are computed from Kohn-Sham orbitals obtained from a prior DFT calculation. We first demonstrate the accuracy of our method on various doped MgH$_2$ supercells. Further, for several transition-metal oxides, we show how the Bader charges obtained with our approach change with the size of the active space and how our method compares to the effect of the Hubbard correction in DFT+U.
            The computational framework behind our many-body calculations, called Dopyqo, is made openly available as a software package.
            \end{minipage}
    \end{center}
    \vspace{1em}
]

\footnotetext[1]{Corresponding author. E-mail address: \href{mailto:erik.schultheis@dlr.de}{erik.schultheis@dlr.de}}

\section{Introduction}
Solving the time-independent Schrödinger equation for molecules and materials gives access to their electronic structure. From this solution, one can calculate relevant electronic properties to characterize quantum systems and to describe their real-world applications. For example, excitation energies, band structures, and optical spectra show their usefulness in many fields such as opto-electronic materials \cite{Meredith2013Electronic, Yu2019Crystal, Ostroverkhova2016Organic}, while the Fermi level, work function and also the ionization energies are of relevance in the characterization of materials to control the transportation of charge carriers \cite{Kahn2016Fermi, Kokil2012Techniques, Poelking2013Characterization, Shirota2007Charge}. Another relevant electronic property is the charge distribution in a material. It is a valuable property for various technologies and materials, such as electrospinning \cite{Collins2012Charge}, conductive polymers \cite{Sessler1997Charge}, 2D materials such as transition metal dichalcogenides \cite{Manzeli20172D}, and doping effects in a material \cite{Hu2019Doping}. A feasible way to evaluate the charge distribution in a material is to compute partial charges such as Löwdin \cite{Loewdin1950OnTheNonOrthogonality} or Bader \cite{Bader1990Atoms} charges.
They can give insight into the charge locality and the nature of the bonding between atoms \cite{Reed1985Natural}.
Löwdin charges are defined by atomic orbitals making them dependent on the chosen basis set \cite{Loewdin1956Natural}. Bader charges require only the real-space electronic density and, therefore, depend on the accuracy of that density \cite{Sanville2007Improved}. Löwdin charges are used for molecules while Bader charges are often used in combination with plane-wave density functional theory~(DFT) codes to describe charge distributions in periodic systems \cite{Fizer2018Benchmark, Thompson2002More, Meng2022APragmatic, Wang2014Modeling}. Bader charge analysis has shown its efficiency in the description of the activation of CO$_2$ by a Ti- and Fe-doped graphene \cite{Giri2024AFirstPrinciples}, and on the role of Mn in decreasing the bandgap in MoS$_2$ \cite{Morales2024Modulating}. Posysaev et al. \cite{Posysaev2019Oxidation} showed that the Bader charge is "a good parameter for predicting oxidation states in most of binary metal oxides". Moreover, Bader charges are also used as charge locality descriptors to evaluate charge transfer properties during electrochemical processes \cite{Kondrakov2017ChargeTransferInduced}. 
The partial charges are dependent on the computational method, and to fully use their potential in the description of material properties, they should be used with other electronic properties such as oscillator strength or dipole moment in order to provide relevant characterization \cite{ Bader1987Properties,Shayeghi2014Optical,Ojanpera2013First}.

As mentioned above, the accuracy of the Bader charges depends strongly on the real-space electronic density which is directly linked to the accuracy of the wave-function solving the Schrödinger equation. However, challenges remain in solving the Schrödinger equation for materials. Indeed, to obtain the electronic wave function or electronic density by solving the time-independent Schrödinger equation, one usually has to approximate the Coulomb interaction between the electrons, since it is very computationally demanding. This approximation makes it difficult to correctly describe materials showing strong electronic correlation. It is expected that such materials deliver relevant industrial value \cite{Cava2021Introduction} but methods like Kohn-Sham density functional theory~(KS-DFT) \cite{Hohenberg1964Inhomogeneous, Kohn1965SelfConsistent} are unable to correctly replicate their properties \cite{Anisimov2010Electronic} since the exchange-correlation (XC) functional, describing the electron-electron interaction, approximates the electronic interaction.
To increase the precision of electronic structure calculations based on DFT, one can extend the XC-functional by a Hubbard-term, i.e., DFT+U \cite{Anisimov1991Band, Anisimov1997FirstPrinciples} or one can use many-body approaches \cite{Ma2020Quantum, Barcza2021DMRG}. Especially many-body calculations are computationally demanding when using methods like coupled-cluster \cite{Coester1958Bound, Bartlett2007CoupledCluster}, configuration-interaction \cite{Roos1972ANewMethod}, or exact diagonalization \cite{Handy1980MultiRoot, Knowles1984ANewDeterminant} to solve the many-body Hamiltonian. In order to reduce the computational cost one can solve the Schrödinger equation within a selected active space built with the orbitals that encapsulate the strong correlations.

Besides algorithms that purely utilize classical computers, there exist various algorithms that either fully run on a quantum computer or are hybrid algorithms that use both classical and quantum resources. Pure quantum algorithms like the quantum phase estimation \cite{Kitaev1995Quantum} use quantum circuits with large gate counts even for small systems, reducing their usefulness for currently available quantum hardware. Variational quantum algorithms \cite{Tilly2022TheVariational} like the variational quantum eigensolver (VQE) \cite{Peruzzo2014AVariational} are hybrid algorithms that are commonly used for solving electronic structure problems. The VQE algorithm was already successfully used to determine the ground state of molecules \cite{OMalley2016Scalable}, periodic systems \cite{Liu2020Simulating} and materials \cite{Ma2020Quantum}.

In this work, we compute Bader charges for various material systems by performing quantum many-body simulations with the VQE algorithm within an active space built from Kohn-Sham orbitals computed with Quantum ESPRESSO \cite{Giannozzi2009Quantum, Giannozzi2017Advanced, Giannozzi2020Quantum}. Moreover, we present a method to compute Bader charges from electronic densities obtained from VQE-optimized wave functions. 
In the theory section, we recall how the matrix elements of the many-body Hamiltonian are computed from the plane-waves based Kohn-Sham orbitals. From the VQE-optimized many-body ground state we compute the charge density and from this the Bader charges on each atom. The results section presents the Bader charges obtained for MgH$_2$ supercells with different elements for the central atom, and various transition metal oxides where DFT calculations typically struggle due to strong correlations. We compare our results to Bader charges calculated with DFT and DFT+U to evaluate how the strong correlations described by our many-body method have an impact on the Bader charges. All many-body calculations are performed with our openly available software package Dopyqo \cite{Dopyqo}.

\section{Theory} \label{sec:theory}
In this section we introduce the theoretical background for our results. We discuss the concepts of DFT \cite{Hohenberg1964Inhomogeneous}, especially its practical implementation by Kohn and Sham \cite{Kohn1965SelfConsistent}. We recall the many-body Hamiltonian and the needed matrix elements used in this work, as well as the frozen core approximation \cite{Yalouz2022Analytical, Rossmannek2021Quantum}, and classical and quantum computing methods to solve the Hamiltonian.

We write vectors in upright bold font ($\brm{v}$), their length in italic font ($v$), sometimes written as $|\brm v|=v$, and operators are denoted with a hat ($\hat{A}$).
A function $\tilde{f}(\cdot)$ is defined as the Fourier transform of the function $f(\cdot)$. We use atomic units throughout this work, which means setting the reduced Planck constant $\hbar$, the electronic charge $e$, the electron mass $m_\mathrm{e}$ and $4\pi\epsilon_0$ to $1$, where $\epsilon_0$ is the vacuum permittivity. We also implicitly use spatial orbitals, when not specified otherwise. Bra-ket notation is used whenever we consider it useful.

\subsection{Density functional theory} \label{sec:dft}

DFT \cite{Hohenberg1964Inhomogeneous} is a reformulation of the many-body time-independent Schrödinger equation in terms of the electronic density. When performing DFT calculations, one typically uses the Kohn-Sham (KS) implementation of DFT \cite{Kohn1965SelfConsistent} where the, formally exact but unknown, exchange-correlation functional is approximated.
During a KS-DFT calculation in a plane-wave basis set representation, each KS orbital is expressed as
\begin{equation} \label{eq:Bloch-exp}
    \braket{\brm{r}|\psi^{(\brm{k})}_t}=\psi_{t}^{(\brm{k})}(\brm{r})=\frac{1}{\sqrt{V}}\sum_{\brm{G},\,G\leq G_\mathrm{cut}}c^{(\brm{k})}_{\brm{G},t}\,e^{i(\brm{k}+\brm{G})\cdot\brm{r}}\,,
\end{equation}
where $\psi_{t}^{(\brm{k})}(\brm{r})$ is the $t$-th KS orbital at the $\brm k$-point in real space. The sum runs over reciprocal lattice vectors $\brm{G}$ with their length $G$ being smaller than a cutoff $G_\mathrm{cut}$ defined by an energy threshold 
$E_\mathrm{cut}=G^2_\mathrm{cut}/2$. In the following, every sum over reciprocal lattice vectors $\brm G$ implicitly has the condition that $G\leq G_\mathrm{cut}$. The complex plane-wave coefficient $c^{(\brm{k})}_{\brm{G},t}$ is associated with the $\brm k$-point on which the $t$-th KS orbital is defined. The $\brm k$-point is a point in the Brillouin zone and, therefore, a vector in reciprocal space. Quantities like the electronic density are defined by an integration over the Brillouin zone. In practice, this integration is approximated by a sum over multiple $\brm k$-points, defined on a k-mesh, where a DFT calculation is performed at every $\brm k$-point. In a self-consistent DFT calculation, the $c^{(\brm{k})}_{\brm{G},t}$ coefficients are optimized to reach the lowest total energy.
It is worth noting here that at the $\brm \Gamma$-point, i.e. $\brm k = \brm \Gamma = \brm 0$, the coefficients $c^{(\brm{k})}_{\brm{G},t}$ can be chosen to be real. 

\subsection{Many-body Hamiltonian and matrix elements}\label{sec:many_body_ham}

In the following, we present the Hamiltonian and its matrix elements as they can be found in the literature \cite{Clinton2024Towards, Yalouz2022Analytical, Rossmannek2021Quantum, Kleinman1982Efficacious, Fraser1996Finite}. We calculate single-electron and two-electron matrix elements to construct a many-body Hamiltonian in the Born-Oppenheimer approximation using the KS orbitals computed from a prior DFT calculation. The single-electron matrix elements $h^{(\brm{k})}_{tu}$ comprise the kinetic energy of the electrons and the pseudopotential while the two-electron matrix elements $h^{(\brm{k})}_{tuvw}$ describe the exact Coulomb interaction. Additionally, the nuclear interaction between all (infinitely many) nuclei and the interaction between each electron and its periodic images contribute constant energies, the nuclear repulsion energy $E_{\mathrm{n\text-n}}$ and the electron self-energy $E_{\mathrm{e\text-self}}$, respectively.
We consider the following electronic structure Hamiltonian $\hat{H}_\mathrm{elec}$ in second quantization:
\begin{equation} \label{eq:hamiltonian-many-body}
    \begin{aligned}
        \hat{H}^{(\brm{k})}_\mathrm{elec} =& \sum_{tu} h^{(\brm{k})}_{tu} \hat{a}^\dagger_t \hat{a}_u + \frac{1}{2} \sum_{tuvw} h^{(\brm{k})}_{tuvw} \hat{a}^\dagger_t \hat{a}^\dagger_u \hat{a}_v \hat{a}_w\\&+ E_{\mathrm{n\text-n}} + E_{\mathrm{e\text-self}}\,
    \end{aligned}
\end{equation}
where $h^{(\brm{k})}_{tu}$ and $h^{(\brm{k})}_{tuvw}$ are the single- and two-electron matrix elements at the $\brm k$-point, respectively. The indices $t$, $u$, $v$, and $w$ denote the indices of the KS band at the $\brm{k}$-point. We define the KS state corresponding to the $t$-th KS band at the $\brm k$-point as
\begin{equation} \label{eq:ks-ket-expansion-in-pw}
    \ket{\psi^{(\brm{k})}_t} = \sum_{\brm{G}} c^{(\brm{k})}_{\brm{G},t} \ket{\brm{k}+\brm{G}}\,,
\end{equation}
which is expanded in a plane-wave basis set using complex coefficients $c^{(\brm{k})}_{\brm{G},t}$ and states $\ket{\brm{k}+\brm{G}}$ where $\brm{G}$ is a reciprocal lattice vector. Since all quantities can be easily defined for arbitrary $\brm{k}$-points, we will omit the $\brm{k}$-point in the following for the sake of clarity, i.e.,
\begin{equation}
    \begin{aligned}
        \ket{\psi^{(\brm{k})}_t} &\to \ket{\psi_t}\,,\\
        c^{(\brm{k})}_{\brm{G},t} &\to c_{\brm{G},t}\,,\\
        \ket{\brm{k}+\brm{G}} &\to \ket{\brm{G}}\,,\\
        h^{(\brm{k})}_{tu} &\to h_{tu}\,,\\
        h^{(\brm{k})}_{tuvw} &\to h_{tuvw}\,.
    \end{aligned}
\end{equation}
Every matrix element in the KS basis can be written as a weighted sum of matrix elements in the plane-wave basis \cite{Clinton2024Towards}. For single-electron matrix elements this means
\begin{equation} \label{eq:single-matrix-elements-ks-to-pw}
    A_{tu}=\bra{\psi_t}\hat{A}\ket{\psi_u}=\sum_{\brm{G}_1,\brm{G}_2} c^\ast_{\brm{G}_1,t} c_{\brm{G}_2,u} \bra{\brm{G}_1}\hat{A}\ket{\brm{G}_2}\,.
\end{equation}
We note that $h_{tu}$ and $h_{tuvw}$ have several symmetries in general and $h_{tuvw}$ has additional symmetries if the wave functions $\psi_t(\brm{r})$ are real \cite{Clinton2024Towards}. 
The symmetries reduce the number of matrix elements that have to be explicitly calculated. We list all relevant symmetries in Appendix~\ref{app:symmeries-matrix-elements}.
 
\subsubsection{Periodicity of the matrix elements}
Let $f(\brm{r})$ be a cell-periodic function with $f(\brm{r}+\brm{T})=f(\brm{r})$, where $\brm{T}$ is a lattice translation vector. Then its Fourier transform \cite{Martin2020Electronic, Clinton2024Towards}
\begin{equation} \label{eq:fourier-transform-of-periodic-function}
    \tilde{f}(\brm{G})=\frac{1}{V} \int_{C} f(\brm{r}) e^{-i\brm{G} \cdot \brm{r}} \mathrm{d}\brm{r}\,,
\end{equation}
where $V$ is the volume of the computational cell $C$ and $\brm{G}$ is a reciprocal lattice vector. So, $\tilde{f}(\brm{p})=0$ if $\brm{p}$ is not a reciprocal lattice vector. We note that for calculating Eq.~\eqref{eq:fourier-transform-of-periodic-function} we only need the function definition of $f(\brm{r})$ inside the computational cell. All potentials used in Sections~\ref{sec:single-electron-matrix-elements}, \ref{sec:nuclear-interaction}, and \ref{sec:two-electron-matrix-elements} are cell-periodic potentials. There, we build a cell-periodic potential $V^{(\brm{T})}(\brm{r})=\sum_{\brm{T}} V(\brm{r}+\brm{T})$ from a non-periodic potential $V(\brm{r})$. The Fourier transform of this cell-periodic potential is
\begin{equation} \label{eq:fourier-transform-of-periodic-function-our-case}
    \begin{aligned}
        \tilde{V}^{(\brm{T})}(\brm{G})&=\frac{1}{V} \int_{C} V^{(\brm{T})}(\brm{r})\,e^{-i\brm{G} \cdot \brm{r}} \mathrm{d}\brm{r}\\
        &= \frac{1}{V} \int_{\mathbb{R}^3} V(\brm{r})\,e^{-i\brm{G} \cdot \brm{r}} \mathrm{d}\brm{r}\,.
    \end{aligned}
\end{equation}
For clarity we will only define the non-periodic potential $V(\brm{r})$, although the actual potential the electrons are experiencing is $V^{(\brm{T})}(\brm{r})$. We then use Eq.~\eqref{eq:fourier-transform-of-periodic-function-our-case} to reflect the cell-periodicity in reciprocal space. 
The periodic potential $V^{(\brm{T})}(\brm{r})$ can be obtained from $\tilde{V}^{(\brm{T})}(\brm{G})$ with \cite{Martin2020Electronic, Clinton2024Towards}
\begin{equation}
    V^{(\brm{T})}(\brm{r}) = \sum_{\brm G} \tilde{V}^{(\brm{T})}(\brm{G}) e^{i\brm G \cdot\brm r}\,.
\end{equation}
 
\subsubsection{Single-electron matrix elements} \label{sec:single-electron-matrix-elements}
We compute matrix elements $h_{tu}$ in the KS basis, i.e.,
\begin{equation}
    h_{tu} = \bra{\psi_t}\hat{T}+\hat{V}_\mathrm{pp}\ket{\psi_u}\,,
\end{equation}
where $\hat{T}$ and $\hat{V}_\mathrm{pp}$ are the kinetic energy and pseudopotential operators, respectively. The kinetic energy operator in the KS basis reads
\begin{equation}
    \bra{\psi_t}\hat{T}\ket{\psi_u} = \frac{1}{2}\bra{\psi_t}\hat{\brm{p}}^2\ket{\psi_u}=\frac{1}{2} \sum_{\brm{G}} c^\ast_{\brm{G},t} c_{\brm{G},u}\,\brm{G}^2\,,
\end{equation}
where we used that the momentum operator $\hat{\brm{p}}$ is diagonal in the plane-wave basis, i.e., $\bra{\brm{G}_1}\hat{\brm{p}}\ket{\brm{G}_2}=\delta_{\brm{G}_1-\brm{G}_2}\,\brm{G}_2$ where $\delta_{\brm{G}}$ is the Kronecker-delta.

To calculate the matrix elements $\bra{\psi_t}\hat{V}_\mathrm{pp}\ket{\psi_u}$ we focus on norm-conserving pseudopotentials in a non-local Kleinman-Bylander form~\cite{Kleinman1982Efficacious}.
Then
\begin{equation} \label{eq:v-pp}
    \begin{aligned}
        \hat{V}_\mathrm{pp} &= \sum_I \left[\hat{V}^{(I)}_\mathrm{loc} + \hat{V}^{(I)}_\mathrm{nl} \right] \\
        &= \sum_I \left[\hat{V}^{(I)}_\mathrm{loc} + \sum_{ijlm} D^{(I)}_{ij} \ket{\beta^{(I)}_{ilm}} \bra{\beta^{(I)}_{jlm}} \right]\,,
    \end{aligned}
\end{equation}
consists of a local ($\hat{V}^{(I)}_\mathrm{loc}$) and a non-local ($\hat{V}^{(I)}_\mathrm{nl}$) part for each nucleus $I$. The states $\ket{\beta^{(I)}_{ilm}}$ are eigenstates of the angular momentum operator.

The local part $\hat{V}^{(I)}_\mathrm{loc}$ is given as a radial-dependent real-space function $V^{(I)}_\mathrm{loc}(r)=\bra{\brm{r}}\hat{V}^{(I)}_\mathrm{loc}\ket{\brm{r}}$, with its Fourier-transform being 
$\bra{\brm{G}_1}\hat{V}^{(I)}_\mathrm{loc}\ket{\brm{G}_2}$. Note that $\bra{\brm{r}}\hat{V}^{(I)}_\mathrm{loc}\ket{\brm{r}^{\prime}}=V^{(I)}_\mathrm{loc}(r)\,\delta(\brm{r}-\brm{r}^{\prime})$, since $\hat{V}^{(I)}_\mathrm{loc}$ is a local potential. 
Introducing an electronic background potential to treat the singularity at $\brm{G}_1=\brm{G}_2$ and subtracting $-Z_I\, \mathrm{erf}(r)/r$ in real-space and adding it again in reciprocal space yields
\begin{equation} \label{eq:v-loc-pw}
    \begin{aligned}
        \bra{\brm{G}_1}\hat{V}^{(I)}_\mathrm{loc}\ket{\brm{G}_2} =&\frac{4\pi}{V} e^{-i(\brm{G}_1-\brm{G}_2)\cdot \brm R_I}\\
        &\times\Bigg[- Z_I \frac{e^{-\left(\brm{G}_1-\brm{G}_2\right)^2/4}}{\left(\brm{G}_1-\brm{G}_2\right)^2}\\
        &+ \int_0^\infty r\,\left(V^{(I)}_\mathrm{loc}(r) + \frac{Z_I\, \mathrm{erf}(r)}{r}\right)\\
        &\phantom{+\int_0^\infty}\ \times\frac{\sin\left(\left|\brm{G}_1-\brm{G}_2\right|r\right)}{\left|\brm{G}_1-\brm{G}_2\right|} \mathrm{d}r\Bigg]\,,
    \end{aligned}
\end{equation}
where $V$ is the volume of the computational cell in real space. $Z_I$ and $\brm{R}_I$ are the charge and the position of the $I$-th nucleus, respectively. The charges $Z_I$ are most often effective core charges since the pseudopotential potentially absorbs some core electrons. 
The zero-momentum part of $V^{(I)}_\mathrm{loc}(r)$ equals the zero-momentum part of its non-Coulombic part which is
\begin{equation} \label{eq:v-loc-pw-zero-mom-lim}
    \left.\bra{\brm{G}_1}\hat{V}^{(I)}_\mathrm{loc}\ket{\brm{G}_2}\right|_{\brm{G}_1=\brm{G}_2}=\frac{4\pi}{V} \int_0^\infty\!\!r^2 \left(V^{(I)}_\mathrm{loc}(r)+\frac{Z_I}{r}\right)\mathrm{d}r\,,
\end{equation}
since the Coulombic part cancels with the electronic background. See Appendix~\ref{app:local-pp-derivation} for derivations of Equations~\eqref{eq:v-loc-pw} and \eqref{eq:v-loc-pw-zero-mom-lim}.

In the non-local part $\hat{V}^{(I)}_\mathrm{nl}$ of the pseudopotential, the states $\ket{\beta^{(I)}_{ilm}}$ are short-ranged in real space:
\begin{equation}
    \braket{\brm{r}|\beta^{(I)}_{ilm}}=\beta^{(I)}_{ilm}(\brm{r})=\beta^{(I)}_{il}(r)\,Y_{lm}(\brm{r})\,,
\end{equation}
with the spherical harmonics $Y_{lm}(\brm{r})$. Using the Rayleigh expansion of plane-waves, the orthonormality of the spherical harmonics, and defining
\begin{equation}
    F^{(I)}_{il}(x)=\int_0^\infty r^2 \beta^{(I)}_{il}(r) j_l(xr) \mathrm{d}r\,,
\end{equation}
where $j_l(x)$ are the spherical Bessel functions, yields
\begin{equation} \label{eq:v-nl-pw}
    \begin{aligned}
        \bra{\brm{G}_1}\hat{V}^{(I)}_\mathrm{nl}\ket{\brm{G}_2}=&\frac{\left(4\pi\right)^2}{V} e^{-i(\brm{G}_1-\brm{G}_2)\cdot \brm R_I} \\
        &\times\sum_{ijlm} D^{(I)}_{ij} Y_{lm}(\brm{G}_1) Y^\ast_{lm}(\brm{G}_2)\\
        &\phantom{\times\sum_{ijlm}\ }\times F^{(I)}_{il}(G_1) F^{(I)}_{jl}(G_2)\,.
    \end{aligned}
\end{equation}
We apply Eq.~\eqref{eq:single-matrix-elements-ks-to-pw} to Eq.~\eqref{eq:v-loc-pw}, Eq.~\eqref{eq:v-loc-pw-zero-mom-lim}, and Eq.~\eqref{eq:v-nl-pw} to obtain $\bra{\psi_t}\hat{V}_\mathrm{pp}\ket{\psi_u}$.

\subsubsection{Nuclear interaction} \label{sec:nuclear-interaction}
The interaction between the nuclei constitutes a constant nuclear repulsion energy $E_{\mathrm{n\text-n}}$ in the Born-Oppenheimer approximation and can be written as
\begin{equation} \label{eq:nuclear-repulsion-energy}
    E_{\mathrm{n\text-n}}=\frac{1}{2}\sum_{I,J}^N \sideset{}{'}\sum_{\brm{T}} \frac{Z_I Z_J}{\left|\brm{R_I}-\brm{R_J}-\brm{T}\right|}\,,
\end{equation}
where we sum over all lattice translation vectors $\brm{T}$ and the prime in the summation indicates that we omit self-interaction terms with $\brm{T}=\bm{0}$ and $I=J$ \cite{Shan2005Gaussian}. $E_{\mathrm{n\text-n}}$ converges slowly when calculated with Equation~\eqref{eq:nuclear-repulsion-energy} but converges rapidly when using the Ewald summation technique \cite{Ewald1921DieBerechnung}. Ewald summation splits Eq.~\eqref{eq:nuclear-repulsion-energy} into a short- and a long-range term which are calculated in real- and reciprocal space, respectively. Then, a self-energy and a charged-system term have to be added to respect the primed summation in Eq.~\eqref{eq:nuclear-repulsion-energy} and the correct zero-momentum term, respectively. Then
\begin{equation} \label{eq:nuclear-repulsion-energy-ewald}
    \begin{aligned}
        E_{\mathrm{n\text-n}}=&\frac{1}{2} \sum_{I,J,\brm{T}} Z_I Z_J \frac{\mathrm{erfc}(\left|\brm{R_I} - \brm{R_J} - \brm{T}\right| \sqrt{\sigma})}{\left|\brm{R_I} - \brm{R_J} - \brm{T}\right|}\\
        &+\frac{2\pi}{V} \sum_{\brm{G} \neq \brm{0}} \sum_{I,J} \frac{Z_I Z_J}{G^2} e^{i \brm{G} \cdot (\brm{R_I} - \brm{R_J}) - \frac{1}{4 \sigma} G^2}\\
        &- \sqrt{\frac{\sigma}{\pi}} \sum_I Z_I^2 - \frac{\pi}{2 V \sigma}  \left(\sum_I Z_I\right)^2
        \,,
    \end{aligned}
\end{equation}
where $\sigma>0$ partitions Eq.~\eqref{eq:nuclear-repulsion-energy} into real and reciprocal space sums, and can be chosen such that the sums converge rapidly in terms of lattice translation vectors $\brm{T}$ and reciprocal lattice vectors $\brm{G}$. In our calculations we choose $\sigma$ such that \cite{Giannozzi2009Quantum, Giannozzi2017Advanced, Giannozzi2020Quantum}
\begin{equation}
    \sum_I 2Z_I \sqrt{\frac\sigma\pi}\,\mathrm{erfc}\left(\sqrt{\frac{\left(4G_\mathrm{cut}\right)^2}{4\sigma}}\right)  \leq 10^{-7}\,,
\end{equation}
where $G_\mathrm{cut}$ is the cutoff energy from Eq.~\eqref{eq:Bloch-exp}. A derivation of Eq.~\eqref{eq:nuclear-repulsion-energy-ewald} can be found in Appendix~\ref{app:nuclear-int-ewald-derivation}.

\subsubsection{Two-electron matrix elements and electron self-energy}
\label{sec:two-electron-matrix-elements}
The potential $V_\text{e-e}(\brm r, \brm r^\prime)$, that defines the periodic electron-electron interaction at position $\brm r$ generated by an electronic point charge at $\brm r^\prime$, is given as
\begin{equation} \label{eq:periodic-e-e-interaction}
    V_\text{e-e}(\brm r, \brm r^\prime) = \sum_{\brm{T}} \frac{1}{|\brm{r}-\brm{r}^\prime-\brm{T}|} - \Phi_b(r)\,,
\end{equation}
where $\brm{T}$ is a lattice translation vector, $\brm r$ and $\brm r^\prime$ lie within the computational cell, 
and $\Phi_b(\brm r)$ is the potential modelling the uniform cancelling background charge of the nuclei. We define $V_\text{e-e}(\brm r, \brm r^\prime)$ only where $\brm r\neq\brm r^\prime$. Equation~\eqref{eq:periodic-e-e-interaction} can be written as an Ewald sum and its Fourier transform $\tilde V_\text{e-e}(\brm G)$ is \cite{Fraser1996Finite}
\begin{equation} \label{eq:periodic-e-e-interaction-fourier-trans}
    \begin{aligned}
        \tilde V_\text{e-e}(\brm G) &= \frac{1}{V}\int_{C} V_\text{e-e}(\brm r, \brm r^\prime) e^{i\brm G \cdot \left(\brm r - \brm r^\prime\right)} \mathrm d \brm r\\
        &= 
        \begin{cases}
            4\pi/\left(VG^2\right), & \text{if } G\neq0 \\
            0, & \text{if } G=0
        \end{cases}\,,
    \end{aligned}
\end{equation}
where $V$ is the volume of the computational cell, $\brm{G}$ is a reciprocal lattice vector, and the potential $\Phi_b(\brm r)$ sets $V_\text{e-e}(\brm 0)$ to zero \cite{Fraser1996Finite}. Note that in the definition of $V_\text{e-e}$ we removed the electron self-interaction $E_{\mathrm{e\text-self}}$, describing the interaction between each electron and its periodic images, which contributes a constant energy and is added back to the Hamiltonian \cite{Fraser1996Finite}:
\begin{equation} \label{eq:electron-self-energy}
    E_{\mathrm{e\text-self}} = N_e \sum_{\brm{T}\neq\bm{0}} \frac{1}{|\brm{T}|}\,,
\end{equation}
where the term with $\brm{T}=\bm{0}$ is unphysical and excluded from the summation. $E_{\mathrm{e\text-self}}$ can be calculated using an Ewald summation (cf.~Eq.~\eqref{eq:nuclear-repulsion-energy-ewald}). We can now define the two-electron matrix elements as
\begin{equation} \label{eq:eri-real-space}
    \begin{aligned}
        h_{tuvw} =& \int\!\!\!\int  \psi^\ast_t(\brm{r}_1)\,\psi^\ast_u(\brm{r}_2)\,V_\text{e-e}(\brm r_1, \brm r_2)\\
        &\hspace{20pt}\times\psi_v(\brm{r}_2)\,\psi_w(\brm{r}_1)\,\mathrm{d}\brm{r}_1 \mathrm{d}\brm{r}_2\,.
    \end{aligned}
\end{equation}
Now we define $\tilde\rho_{uv}(\brm{G})$ as the Fourier transform of pair-densities $\rho_{uv}(\brm{r})=\psi^\ast_u(\brm{r}) \psi_v(\brm{r})$ which is equivalent to a convolution in reciprocal space, i.e., $\tilde\rho_{uv}(\brm{G})=\big(\tilde\psi_u(-\brm{G})\big)^\ast * \tilde\psi_v(\brm{G})$, where $*$ is the convolution operation. 
We note that the Fourier transform of the pair density satisfies $\tilde\rho_{il}(\brm{G}) = \tilde\rho^\ast_{li}(-\brm{G})$.
Using Eq.~\eqref{eq:periodic-e-e-interaction-fourier-trans} and Eq.~\eqref{eq:fourier-transform-of-periodic-function} we find
\begin{equation} \label{eq:eri-reciprocal-space}
        h_{tuvw} = 4\pi V \sum_{\brm{G}, \brm{G}\neq\bm{0}} \frac{\tilde\rho^{\ast }_{wt}(\brm{G}) \tilde\rho_{uv}(\brm{G})}{G^2}\,.
\end{equation}
When performing calculations using multiple $\brm k$-points, it is important to consider that the two-electron matrix elements couple different $\brm k$-points based on crystal momentum conservation \cite{Sun2023Exact}.

\subsubsection{Frozen core approximation} \label{sec:frozen_core_approx}
We use the frozen core approximation \cite{Yalouz2022Analytical,Rossmannek2021Quantum} to perform active-space calculations involving only some of the electrons in a limited number of orbitals. The complete system involving all electrons is split into two parts, an active space and an environment. Only electrons in the active space are simulated explicitly while all other electrons in the environment are modelled by modifying the matrix elements of the Hamiltonian. This reduces the computational effort since the number of electronic configurations now depends only on the size of the active space. 

In the frozen core approximation the orbitals in the environment are assumed to be fully-occupied and electrons cannot be transferred between the environment and the active space. This results in an energy offset, the frozen core energy $E_\mathrm{frozen}$, and an effective single-electron potential with matrix elements $g_{tu}$. In the following we use the indices $\{a,b\}$ for orbitals in the environment and $\{t,u,v,w\}$ for orbitals in the active space. The frozen core energy equals
\begin{equation} \label{eq:frozen-core-energy}
    E_\mathrm{frozen} = 2 \sum_a h_{aa} + \sum_{ab} \left(2h_{abba}-h_{abab}\right)\,,
\end{equation}
where we assume that the spin-orbitals are identical for both spins, and $a$ and $b$ run over all fully-occupied spatial orbitals in the environment. The matrix elements of the effective single-electron potential equal
\begin{equation} \label{eq:frozen-core-potential}
    g_{tu} = \sum_{a} \left(2 h_{taau} - h_{taua}\right)\,,
\end{equation}
where the $t$ and $u$ index spatial orbitals in the active space. See Appendix~\ref{app:frozen-core-spin} for a derivation and a spin-resolved definition of the frozen core approximation.

With this the active-space Hamiltonian within the frozen core approximation in second quantization reads
\begin{equation}
    \label{eq:hamiltonian-many-body-frozen-core}
    \begin{aligned}
        \hat{H}_\mathrm{elec} =& \sum_{tu} \left(h_{tu} + g_{tu}\right) \hat{a}^\dagger_t \hat{a}_u\\&+ \frac{1}{2} \sum_{tuvw} h_{tuvw} \hat{a}^\dagger_t \hat{a}^\dagger_u \hat{a}_v \hat{a}_w\\&+ E_{\mathrm{n\text-n}} + E_{\mathrm{e\text-self}} + E_\mathrm{frozen}\,.
    \end{aligned}
\end{equation}
It is worth noting that there are alternatives to the frozen core approximation that go beyond the HF-like mean-field description for the electrons not in the active space. For example, the WF-in-DFT approach \cite{Manby2012ASimple, Lee2019ProjectionBased} embeds the active space into a potential generated by the DFT density. While this approach is more involved than the frozen core approximation it reuses the prior DFT calculation resulting in a potentially better description of the electronic structure. For the present work we focus on the frozen core approximation because of its simplicity while future work explores other embedding techniques.
\\
We note that the Kohn–Sham orbitals are used only as a single-particle basis. Neither the Kohn–Sham eigenvalues nor the DFT Hartree or exchange-correlation energy enters the active-space Hamiltonian. Instead, the one- and two-electron matrix elements are evaluated from the kinetic-energy and pseudopotential operators and the exact Coulomb interaction. After freezing the inactive orbitals, their contribution is represented by Hartree–Fock-like Coulomb and exchange terms. Thus, no DFT exchange-correlation contribution is combined with the energy from solving the many-body Hamiltonian, and no separate double-counting correction is introduced.

\section{Variational quantum eigensolver} \label{sec:vqe}
The VQE \cite{Peruzzo2014AVariational} is a hybrid quantum-classical algorithm that utilizes a quantum computer to prepare a parameterized quantum state which is classically optimized with regard to its energy given a Hamiltonian. The state preparation is implemented as a unitary operation, often called ansatz, acting on qubits of the quantum computer.
We use a VQE to find the ground state of the Hamiltonian defined in Eq.~\eqref{eq:hamiltonian-many-body-frozen-core}.

We use the VQE in combination with a particle-number preserving unitary coupled cluster with single and double excitations (UCCSD) ansatz \cite{Bartlett1989Alternative, Taube2006New} to prepare a parameterized quantum state. The UCCSD ansatz $\hat U(\bm{\theta})$ is composed of fermionic single excitations
\begin{equation}
    \hat G^{p}_{i} = i \left(\hat a^\dagger_{p} \hat a_{i}-\mathrm{h.c.}\right)\,,
\end{equation}
and fermionic double excitations
\begin{equation}
    \hat G^{pq}_{ij} = i \left(\hat a^\dagger_{p} \hat a^\dagger_{q} \hat a_{j} \hat a_{i}-\mathrm{h.c.}\right)\,.
\end{equation}
A first-order Trotter-Suzuki \cite{Trotter1959OnTheProduct, Suzuki1976Generalized} decomposition yields
\begin{equation} \label{eq:uccsd-trotterized}
    \hat U(\bm{\theta}) = \prod_{p, i} \exp\left(-i\theta^{p}_{i} \hat G^{p}_{i} \right) \prod_{p,q,i,j} \exp\left(-i\theta^{pq}_{ij} \hat G^{pq}_{ij} \right)\,,
\end{equation}
where the indices $i$, $j$ and $p$, $q$ go over all occupied and virtual spin orbitals, respectively. We note that $\hat U(\bm{\theta})$ is applied in the quantum circuit after a reference determinant is prepared. 
In this work, we use as reference determinant the Slater determinant where the first $N/2$ active spatial orbitals are occupied and the rest are unoccupied, where $N$ is the number of active electrons. We note that this is not a Hartree-Fock state since no Hartree-Fock calculation was performed.

We note that the computational framework we present here is not limited to the VQE algorithm. Indeed, any quantum algorithm can be used that is able to prepare the ground state of a Hamiltonian. We choose VQE, since it is an algorithm that can be run on current quantum hardware and is a promising algorithm to initialize other quantum algorithms like quantum phase estimation \cite{Kitaev1995Quantum}. To showcase the computational framework and its application to quantum computing, we restrict ourselves to VQE while future work will involve investigating alternatives to VQE or different ansätze and optimization strategies within VQE. We note that simulation of VQE using a UCCSD ansatz on classical computers is limited by the computational cost of the optimization loop, which increases with the number of parameters and iterations required to converge to a solution. This substantially limits the size of the active spaces we are able to simulate.

\section{Computational cost and limitations} \label{sec:limitations}


The computational cost of KS-DFT is dominated by the diagonalization of the Kohn-Sham Hamiltonian, which scales as $\mathcal O(N_\mathrm{pw}^3)$, where $N_\mathrm{pw}$ is the number of basis functions used in KS-DFT, plane-waves in our case \cite{Mohr2014Daubechies}. 
DFT+U also formally scales as $\mathcal O(N_\mathrm{pw}^3)$ although practical calculations can take much longer than KS-DFT calculations of the same system \cite{CapdevilaCortada2016Performance}.
On the other hand, the computational cost of solving our many-body Hamiltonian depends on the used method. Classical correlated methods like coupled-cluster or configuration-interaction scale with $\mathcal O(N_b^6)$ when including single and double excitations \cite{Cremer2013FromConfiguration, Bartlett2007CoupledCluster}. 
Here, $N_b$ is the number of basis functions spanning the active space, which are KS-orbitals in our case. Exact solvers like the complete active space configuration interaction (CASCI) solver scale exponentially in $N_b$ since they resort to exact diagonalization of the many-body Hamiltonian in the full Hilbert space of $\mathcal O(2^{N_b})$ determinants.
VQE uses $N_b$ qubits, while the circuit depth of, e.g., a UCCSD ansatz scales with $\mathcal O(N_b^5)$ \cite{Fan2023CircuitDepth}. Also, many-body electronic structure Hamiltonians have $\mathcal O(N_b^4)$ matrix elements, compared to $\mathcal O(N_\mathrm{pw}^2)$ matrix elements in KS-DFT.
Therefore, the computational cost of our many-body approach is significantly larger than that of KS-DFT, but the Hamiltonian can more accurately describe the electron-electron interaction as we detail in the following.

In KS-DFT, the electrons are usually over-delocalized due to the unphysical electron self-interaction coming from the Hartree term, which is not exactly cancelled by an approximate exchange correlation potential \cite{Himmetoglu2013HubbardCorrected}. Thus, energy gaps, magnetic moments, charge distribution, insulator or metallic behavior are not well described for some systems \cite{Malet2014Exchange}. 
The introduction of a Hubbard U energy correction in DFT+U leads to localization of selected states, which are usually the $d$ and $f$ orbitals \cite{Lichtenstein1995Density}. Therefore, DFT+U can remedy the over-delocalization of electrons in KS-DFT \cite{Himmetoglu2013HubbardCorrected}.
Since our many-body approach uses exact exchange by construction, we do not encounter the same self-interaction errors as KS-DFT and therefore avoid over-delocalization of electrons. This approach is similar to Hartree-Fock~(HF) based methods for simulating materials, where electrons are usually over-localized \cite{MoriSanchez2008Localization, Li2017Localized}.
We use post-HF methods to solve our many-body Hamiltonian which correct for this over-localization. Therefore, within the active space, the localization of the electrons is described without error. On the other hand, all electrons in the frozen core are described on a HF-like mean-field level which makes them prone to over-localization. This can lead to degraded quality compared to KS-DFT if the electrons in the frozen core contribute significantly to the electronic structure of the material. Therefore, the active space has to be chosen carefully. In general, it is difficult to formally compare the effects of over-(de)localization on the DFT level and in our many-body Hamiltonian. 
But, since electron over-(de)localization manifests in the charge density, we think that the charge density or quantities derived from it, as the Bader charge, are a suitable indicator to what extent the different methods describe the electronic structure of various systems.

\section{Bader charges}

Bader charges are commonly used in theoretical and experimental chemistry to obtain chemical insight into particular physical behavior of materials. They have been introduced first by Richard Bader \cite{Bader1990Atoms} to quantify charges in molecules. The method is based on the discretization of the electronic charge density, making it independent of the chosen basis set. Thus, Bader charge analysis is equally defined for molecules described with atomic orbitals as for periodic cells described with plane-waves. The accuracy of the Bader charges depends only on the precision of the electronic structure calculation which in turn relies on the accuracy of the basis set.

The Bader method is solely based on the charge density $\rho(\brm r)$ which is a central quantity in DFT and can be calculated from a many-body wave function. To compute Bader charges, first, the real-space is divided into Bader regions bound by zero-flux surfaces running through minima of the charge density. Integrating the charge density within one Bader region yields the corresponding Bader charge of that region. If a Bader region encapsulates one nucleus the Bader charge of that region is assigned to the atom of that nucleus.
The charge densities in our calculations are defined on a grid. To obtain Bader charges from such a charge density, specialized algorithms exist that carefully determine and assign each grid point to a Bader region \cite{Tang2009AGridBased, Sanville2007Improved, Henkelman2006AFast, Yu2011Accurate}. 
The Bader charge $Q_I$ associated with the $I$-th atom reads
\begin{equation}
    Q_I=\int_{B_I}\rho(\brm{r})\,\mathrm{d}^3\brm{r}\,,
\end{equation}
where we integrate the real-space charge density $\rho(\brm{r})$ over the Bader region $B_I$.

In general, the charge density is expressed as 
\begin{equation}
    \rho(\brm r) = \sum_{tu} \gamma_{tu} \psi^\ast_t(\brm r) \psi_u(\brm r)
\end{equation}
where $\gamma_{tu} = \bra\Psi \hat{a}^\dagger_t \hat{a}_u \ket{\Psi}$ is the one-electron reduced density matrix. For DFT calculations $\gamma_{tu}$ is diagonal and $\gamma_{tt}=f_t$ where $f_t$ is the occupancy of the $t$-th KS orbital $\psi_t(\brm{r})$.
Therefore, in DFT, the charge density can be expressed as
\begin{equation}
    \rho(\brm{r})=\sum_t f_t \left|\psi_t(\brm{r})\right|^2\,.
\end{equation}
When running a quantum simulation with VQE, using the Jordan-Wigner fermion-to-qubit mapping \cite{Jordan1928Ueber} and assuming a real-valued wave function we can obtain the matrix elements $\gamma_{tu}$ by estimating
{
\begin{equation}
    \gamma_{tu}=\frac{1}{4}\bra{\Psi_{\mathrm{VQE}}} \left( \prod_{t+1}^{u-1} \hat Z_k \right) (\hat X_t \hat X_u + \hat Y_t \hat Y_u) \ket{\Psi_{\mathrm{VQE}}}\,,
\end{equation}
}
\noindent for off-diagonal elements, where we assumed $t<u$, and
\begin{equation}
    \gamma_{tu}=\frac{1}{2}\bra{\Psi_{\mathrm{VQE}}}\hat I_t - \hat Z_t\ket{\Psi_{\mathrm{VQE}}}\,,
\end{equation}
for diagonal elements, where $\hat I_t$, $\hat X_t$, $\hat Y_t$, $\hat Z_t$ are the identity-, Pauli-X, -Y, and -Z operators acting on the qubit representing the $t$-th orbital, respectively.

\section{Results} \label{sec:results}

The results we show in this section are obtained for various crystal structures and are compared to reference data. We compare to three references and use several crystal structures therein.

\begin{figure}[t]
    \centering
    \includegraphics[width=0.8\columnwidth]{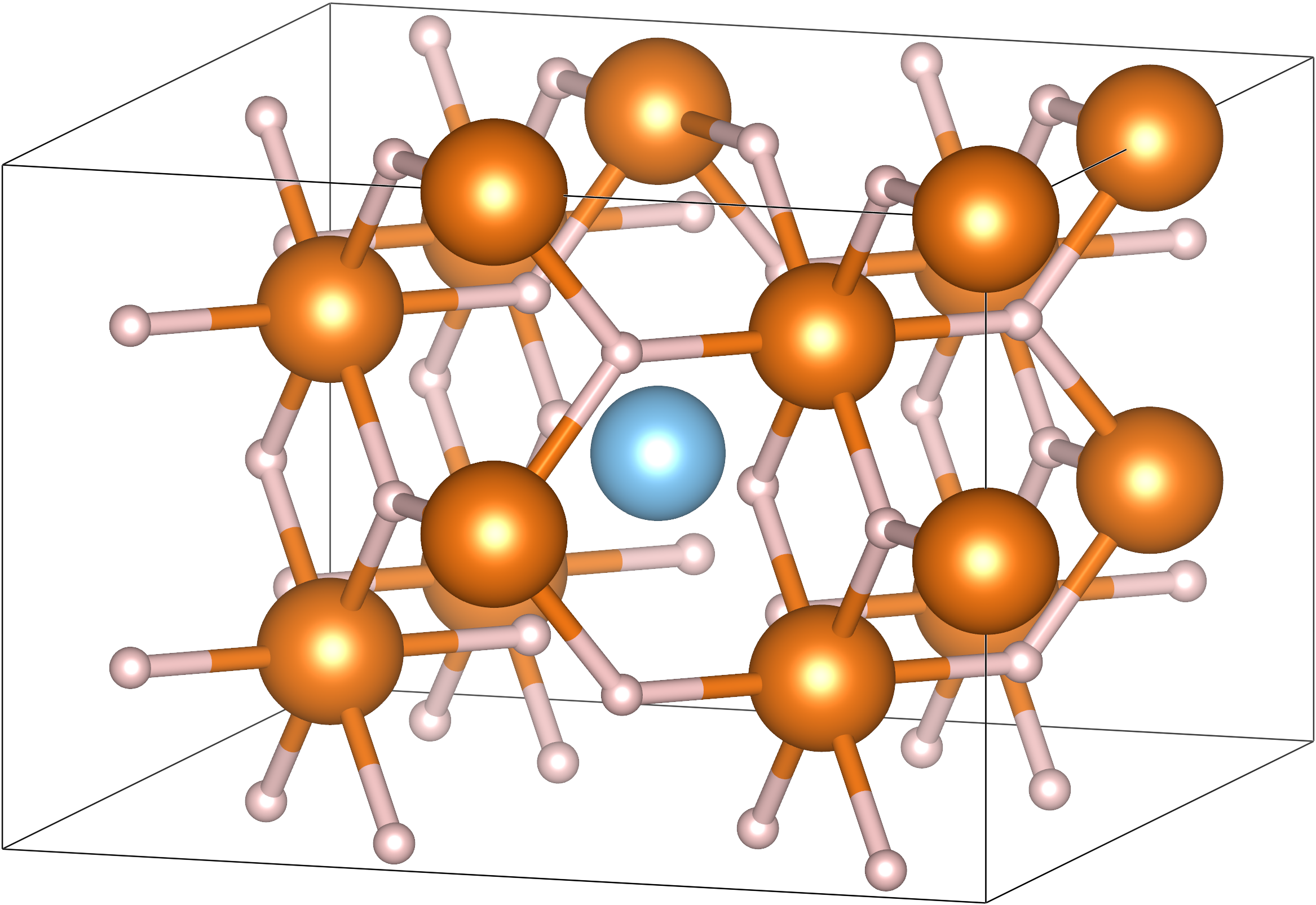}
    \caption{The computational cell used for crystal structures taken from \cite{PaskasMamula2014Electronic} representing a $2\times 2\times 2$ supercell of MgH$_2$. The orange spheres represent an Mg atom, the white spheres are hydrogen atoms and the central light blue sphere is either a Mg atom or a TM atom (Ti, Cr, Fe, Ni, Zn).}
    \label{fig:MgH2_w_doppants}
\end{figure}

Six crystal structures are taken from \cite{PaskasMamula2014Electronic} and they are each described by supercells. The first crystal is a $2\times 2\times 2$ supercell of MgH$_2$ and is built from a single lattice cell where MgH$_2$ has a rutile structure and has crystallized in the tetragonal $\mathrm{P}4_2/\mathrm{mnm}$ space group. The five other crystal structures are built by replacing the central Mg atom of the MgH$_2$ supercell with a transition metal (TM) -- Ti, Cr, Fe, Ni, or Zn. The structures of this set are shown in Fig.~\ref{fig:MgH2_w_doppants}.

One crystal structure is taken from \cite{Lu2018Unraveling}. CrO$_2$ has a rutile structure and has crystallized in the tetragonal $\mathrm{P}4_2/\mathrm{mnm}$ space group.

Five crystal structures are taken from \cite{Choudhuri2020Calculating}. Four transition metal dioxides ($\mathrm{TMO}_2$), with $\mathrm{TM}\in\{\mathrm{Ti},\,\mathrm{Mo},\,\mathrm{Ru},\,\mathrm{Rh}\}$, which are rutile structures and have crystallized in the tetragonal $\mathrm{P}4_2/\mathrm{mnm}$ space group, and one transition metal disulfide ($\mathrm{TMS}_2$), with $\mathrm{TM}=\mathrm{Ti}$.
    
\subsection{Computational methods}
We perform all DFT calculations with the Quantum ESPRESSO software \cite{Giannozzi2009Quantum, Giannozzi2017Advanced, Giannozzi2020Quantum} based on plane-waves and pseudopotentials. We use norm-conserving (NC) pseudopotentials from the SG15 optimized norm-conserving Vanderbilt (ONCV) pseudopotentials which are generated with the ONCVPSP (optimized norm-conserving Vanderbilt pseudopotential) code in a scalar-relativistic version \cite{Hamann2013Optimized}. 
More information on the pseudopotentials is given in Appendix~\ref{app:electrons-in-pp}. We use the generalized gradient approximation of Perdew, Burke, and Ernzerhof (PBE) \cite{Perdew1996Generalized} for our DFT calculations unless stated otherwise. VESTA \cite{Momma2011VESTA} is used for crystal visualization.

The Bader charges are calculated with standard DFT with NC pseudopotential and also with our software package Dopyqo \cite{Dopyqo}. They are compared to reference Bader charges \cite{PaskasMamula2014Electronic,Lu2018Unraveling,Choudhuri2020Calculating} obtained with either standard DFT and DFT+U. The computational parameters reported in Ref.~\cite{PaskasMamula2014Electronic} have been used for the MgH$_2$ systems. Spin-unpolarized calculations are performed for the crystal structures from \cite{PaskasMamula2014Electronic}. A k-mesh of $3\times 3\times 4$ and an energy cutoff value of $816\,\mathrm{eV}$ for the plane-wave expansion of the wave function are used. The atomic positions and cell parameters of the $\mathrm{MgH}_2$ supercell were relaxed, then we replaced the central Mg atom by a TM and only the atomic positions were relaxed again.

For CrO$_2$ and the structures taken from \cite{Choudhuri2020Calculating}, we keep the lattice vectors of the starting cells fixed and relax only the atomic positions in spin-unpolarized calculations with a k-mesh of $12\times 12\times 12$. In Ref.~\cite{Lu2018Unraveling}, CrO$_2$ has been simulated using a spin-polarized calculation. A convergence criterion of $10^{-5}\,\mathrm{eV}$ per atom and of $2\times10^{-4}\,\mathrm{eV}/$\AA\ for the DFT energies and forces are used, respectively. For TiS$_2$, we use a cell containing two Titanium and four Sulfur atoms. We use energy cutoffs of $100\,\mathrm{Ry}$ for CrO$_2$ and TiO$_2$, $120\,\mathrm{Ry}$ for MoO$_2$, RhO$_2$, and RuO$_2$, and $80\,\mathrm{Ry}$ for TiS$_2$ for the plane-wave expansion of the wave function, and cutoffs of $800\,\mathrm{Ry}$ for CrO$_2$, RhO$_2$, RuO$_2$, and TiO$_2$, $1000\,\mathrm{Ry}$ for MoO$_2$, and $600\,\mathrm{Ry}$ for TiS$_2$ for the charge density. On the relaxed geometry, we perform a spin-unpolarized and a spin-polarized DFT-NC calculation at the $\brm\Gamma$-point. Both the spin-unpolarized and the spin-polarized DFT-NC calculations at the $\brm\Gamma$-point use the same relaxed geometry and cutoffs. It is important to highlight at this point that every geometry relaxation is performed with a k-mesh while all Bader charges and Dopyqo calculations are based on a DFT calculation performed at the $\brm\Gamma$-point with a relaxed geometry coming from a k-mesh calculation. Unless stated otherwise, DFT calculations are done with NC pseudopotentials which is different from Refs.~\cite{PaskasMamula2014Electronic, Lu2018Unraveling, Choudhuri2020Calculating}, where projector augmented wave~(PAW) pseudopotentials are used.
For all six transition-metal dioxide and disulfide systems, the computational cells contain six atoms: two transition-metal atoms and four oxygen or sulfur atoms.

The active spaces for our many-body Hamiltonian (Eq.~\eqref{eq:hamiltonian-many-body-frozen-core}) are chosen based on the DFT energies of the KS orbitals calculated by Quantum ESPRESSO. The active spaces are selected by considering orbitals around the Fermi energy. We are careful to involve all orbitals degenerate in energy. For the MgH$_2$ systems, no more than eleven orbitals are involved due to high computational cost when running VQE calculations. We do not include orbitals in the active space that have a DFT energy farther than $5$\,eV from the Fermi energy.
For the TMO systems, we report CASCI results for various active spaces around the Fermi energy with up to $13$ spatial orbitals, resulting in between $23$ and $33$ active spaces per system. The KS energies of the TMO systems at the $\brm\Gamma$-point are shown in Fig.~\ref{fig:ks_energies} in Appendix~\ref{app:active_space_scan}. For each TMO system, we additionally select one representative active space with ten spatial orbitals for the VQE calculations. Among all computed active spaces with ten spatial orbitals, we consider those whose CASCI BEC lies within the range of the CASCI BECs of the active spaces with $11$ to $13$ spatial orbitals, up to a tolerance of $0.005\,\mathrm{e}$. From these, we choose the active space whose number of active electrons is closest to half filling, i.e., closest to the number of spatial orbitals.

For all many-body calculations we use the Dopyqo code \cite{Dopyqo}. We obtain the many-body Hamiltonian (Eq.~\eqref{eq:hamiltonian-many-body-frozen-core}) by calculating the needed matrix elements from the Quantum ESPRESSO output files. We use TenCirChem \cite{Li2023TenCirChem} with its PySCF \cite{Sun2020PySCF} backend to estimate the energy of the UCCSD ansatz using a ground state simulation. The parameters of the ansatz are optimized with the L-BFGS optimizer \cite{Broyden1970TheConvergence, Fletcher1970ANewApproach, Goldfarb1970AFamily, Shanno1970Conditioning, Liu1989OnTheLimited} implemented in SciPy \cite{Virtanen2020Scipy}. In the UCCSD ansatz we start from the Slater determinant where the first $N/2$ active orbitals are occupied and the rest are unoccupied, where $N$ is the number of active electrons. Then all double and single excitations are applied.
This UCCSD ansatz is repeated one time for the CrO$_2$ system to increase the expressivity of the ansatz and reduce the Trotterization error, in order to obtain converged VQE energies. With a single UCCSD layer, the VQE optimization for \ce{CrO_2} converges to a state with $\braket{\hat S^2} = 0.02$ and an energy $22\,\mathrm{mHa}$ above the CASCI energy, whereas two layers yield $\braket{\hat S^2} = 6.00$, corresponding to the $S=2$ CASCI ground state, and an energy within $0.17\,\mathrm{mHa}$ of the CASCI energy. For all other TMO systems, a single UCCSD layer is used.

Complete Active Space Configuration Interaction (CASCI) calculations were carried out using the PySCF~\cite{Sun2020PySCF} package, employing the same active spaces as those defined for the VQE. For the TMO systems, CASCI calculations are additionally carried out for various other active spaces as described above.
We note that both the VQE and CASCI calculations conserve the spin-projection $S_z$ but not the total spin $S$.

The Bader charge analysis code \cite{Arnaldsson2023BaderCode} reads the charge density calculated from the DFT and VQE solution and assigns a Bader charge to each atom in the computational cell. The charge density is represented on a real-space
\makebox[\linewidth][s]{grid of same size as used in the Quantum ESPRESSO}
code~\cite{Giannozzi2009Quantum, Giannozzi2017Advanced, Giannozzi2020Quantum}. 
We do not report the Bader charges directly but use Bader excess charges (BECs) \cite{PaskasMamula2014Electronic}. A BEC is defined as the difference between the computed Bader charge $Q_I$ and the nuclear charge $Z_I$ of the $I$-th atom (as introduced in Sec.~\ref{sec:theory}).

\subsection{Bader Excess Charges in Magnesium Systems and Transition Metal Oxides}
In this section we present the relative effect of different types of computational methods on the atomic BECs in comparison to reference data \cite{PaskasMamula2014Electronic, Lu2018Unraveling, Choudhuri2020Calculating}.
Figure~\ref{fig:BC_MgH2} presents the BECs computed for the \ce{MgH_2} systems, across all methods employed in this work -- DFT, CASCI, and VQE -- and compares them against the reference values of Ref.~\cite{PaskasMamula2014Electronic}. Since the reference does not employ any method specifically designed to improve the description of electronic correlation, it serves here as a baseline against which our many-body approach can be validated. Figure~\ref{fig:bader_TMO} presents the BECs computed for the strongly correlated TMO systems. The results obtained with DFT, CASCI, and VQE are compared against reference values from Refs.~\cite{Lu2018Unraveling, Choudhuri2020Calculating}, allowing us to assess how the active-space many-body treatment changes the charge distribution relative to DFT-NC and the reference values.
Unless stated otherwise, we label any DFT calculation at the $\brm\Gamma$-point using a NC pseudopotential with "DFT-NC". We label any VQE calculation performed on top of a DFT-NC calculation with "VQE". All Complete Active Space Configuration Interaction results are labelled CASCI in all figures. The active spaces used for the CASCI and VQE calculations are specified in Fig.~\ref{fig:BC_MgH2} and Figure~\ref{fig:bader_TMO}, and are displayed along the x-axis of each figure.

\subsubsection{\texorpdfstring{MgH$_2$}{MgH2} systems} \label{sec:bec_mgh2}
\begin{figure}[t]
    \centering
    \includegraphics[width=\columnwidth]{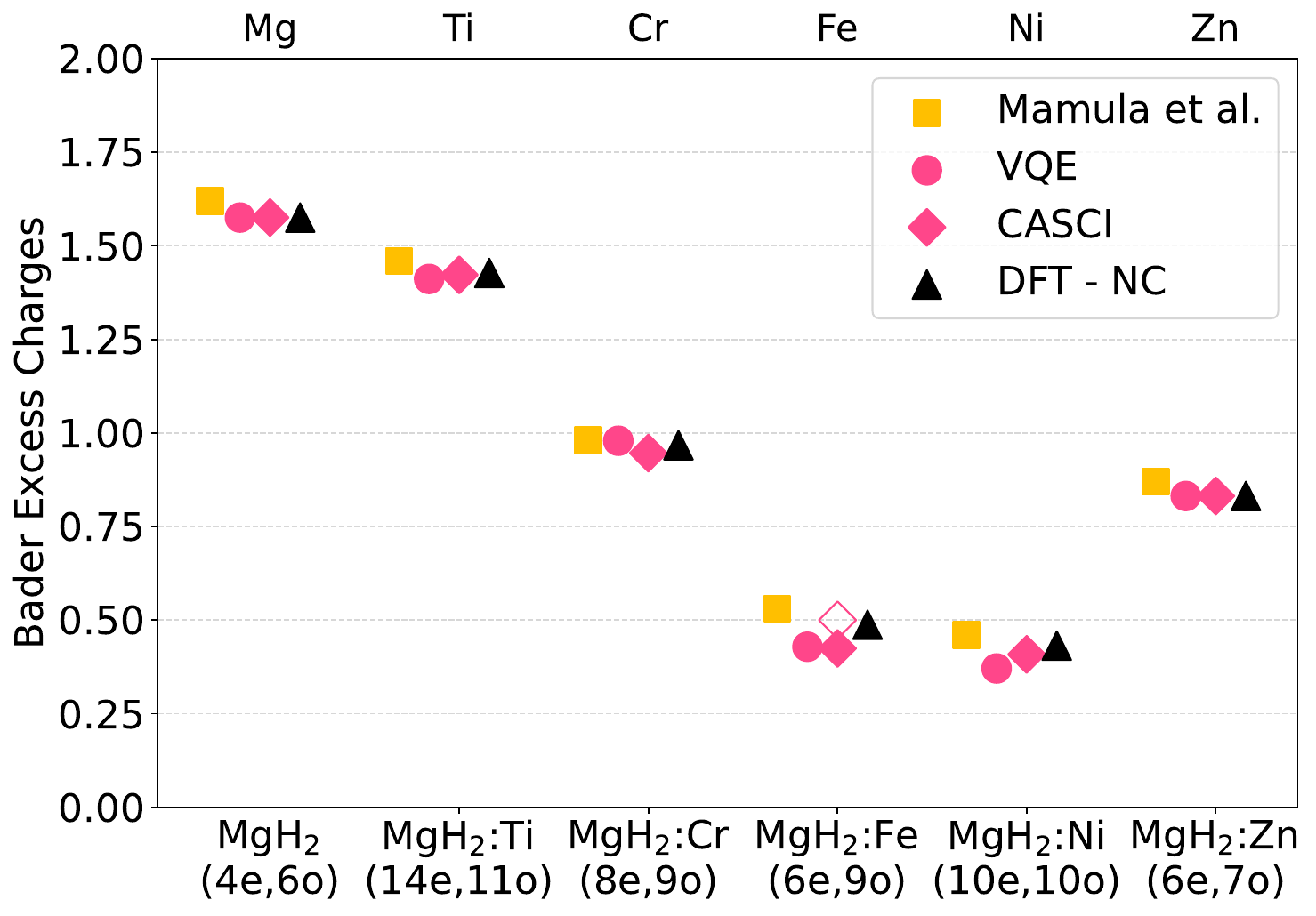}
    \caption{BECs on the central atom of the $2\times 2\times 2$ supercell of MgH$_2$. The yellow squares are the reference data points~\cite{PaskasMamula2014Electronic}, black triangles are DFT-NC results, pink diamonds are CASCI results and the pink circles are VQE results. The pink-outlined diamond corresponds to a larger active space of (18e, 13o).}
    \label{fig:BC_MgH2}
\end{figure}

The BECs computed for the dopant atoms (Mg, Ti, Cr, Fe, Ni and Zn) across all MgH$_2$ systems are reported in Figure~\ref{fig:BC_MgH2} and compared against the reference values from
\noindent Ref.~\cite{PaskasMamula2014Electronic}, which were obtained using spin-unpolarized DFT calculations with PAW pseudopotentials. 
For MgH$_2$,  MgH$_2$:Ti, MgH$_2$:Cr and MgH$_2$:Zn, our DFT values are in close agreement with the reference. For instance, the BEC on Ti in  MgH$_2$:Ti is $1.43\,\mathrm{e}$ (DFT) compared to the reference value of $1.46\,\mathrm{e}$, while for Zn in MgH$_2$:Zn all methods consistently yield $0.83\,\mathrm{e}$.
The CASCI and VQE charges for these systems are equally in good agreement with the reference. Since the reference calculations were performed exclusively at the DFT level, no method capable of treating strong electron correlation was employed therein. Moreover, DFT is known to be inadequate for systems exhibiting strong correlation. The many-body methods CASCI and VQE would in principle be expected to yield corrections relative to the single-reference descriptions. With the exception of MgH$_2$:Fe and MgH$_2$:Ni, the absence of any such correction in the BEC here is consistent with the interpretation that these systems do not exhibit significant strong correlation, even in the presence of transition metals. In this regime, the agreement between the CASCI and VQE charges and the reference therefore serves as a validation of the present computational framework, confirming that the active-space construction and the many-body treatment are internally consistent, even if not strictly required.

A similar conclusion holds broadly for MgH$_2$:Fe and MgH$_2$:Ni, though several points require further discussion.
The CASCI BEC for Ni ($0.41\,\mathrm{e}$) remains close to the reference value ($0.46\,\mathrm{e}$), whereas the VQE result ($0.37\,\mathrm{e}$) deviates more noticeably. This deviation is attributed to incomplete convergence of the VQE optimization; a fully converged VQE calculation is expected to recover the CASCI result, as both methods target the same correlated ground state within the active space.
For MgH$_2$:Fe, the calculations presented here employ an active space of (6e, 9o). Within this active space, VQE and CASCI yield identical BECs ($0.42\,\mathrm{e}$), yet both underestimate the reference value ($0.53\,\mathrm{e}$). This discrepancy is attributed to the limited size of the active space, which does not capture the full correlation relevant to the charge distribution on the dopant (Fe). In the VQE and CASCI calculations, the frozen core electrons are described at a HF-like mean field level, whereas DFT captures their correlation via the exchange-correlation functional. This difference is significant, explaining why DFT is in closer agreement with the reference than CASCI or VQE for this active space. Expanding the active space to (18e, 13o), the CASCI BEC converges towards the reference (shown as a pink-outlined diamond in Figure~\ref{fig:BC_MgH2}), demonstrating the sensitivity of the charge distribution to the active space size in this system. A VQE calculation with the enlarged active space is expected to yield a consistent result upon convergence, but a simulation could not be performed due to prohibitively large computational time. 

For the \ce{MgH_2}:\ce{Cr} and \ce{MgH_2}:\ce{Ni} systems, the VQE optimization did not reach full convergence. This may be attributed to the optimizer becoming trapped in a local minimum, or to the absence of triple excitations in the ansatz. The Trotterization error can be ruled out as a contributing factor, as the BECs remain invariant with respect to the number of Trotter steps.

\subsubsection{Transition Metal Oxides}

\begin{figure}[t]
    \centering
    \includegraphics[width=\columnwidth]{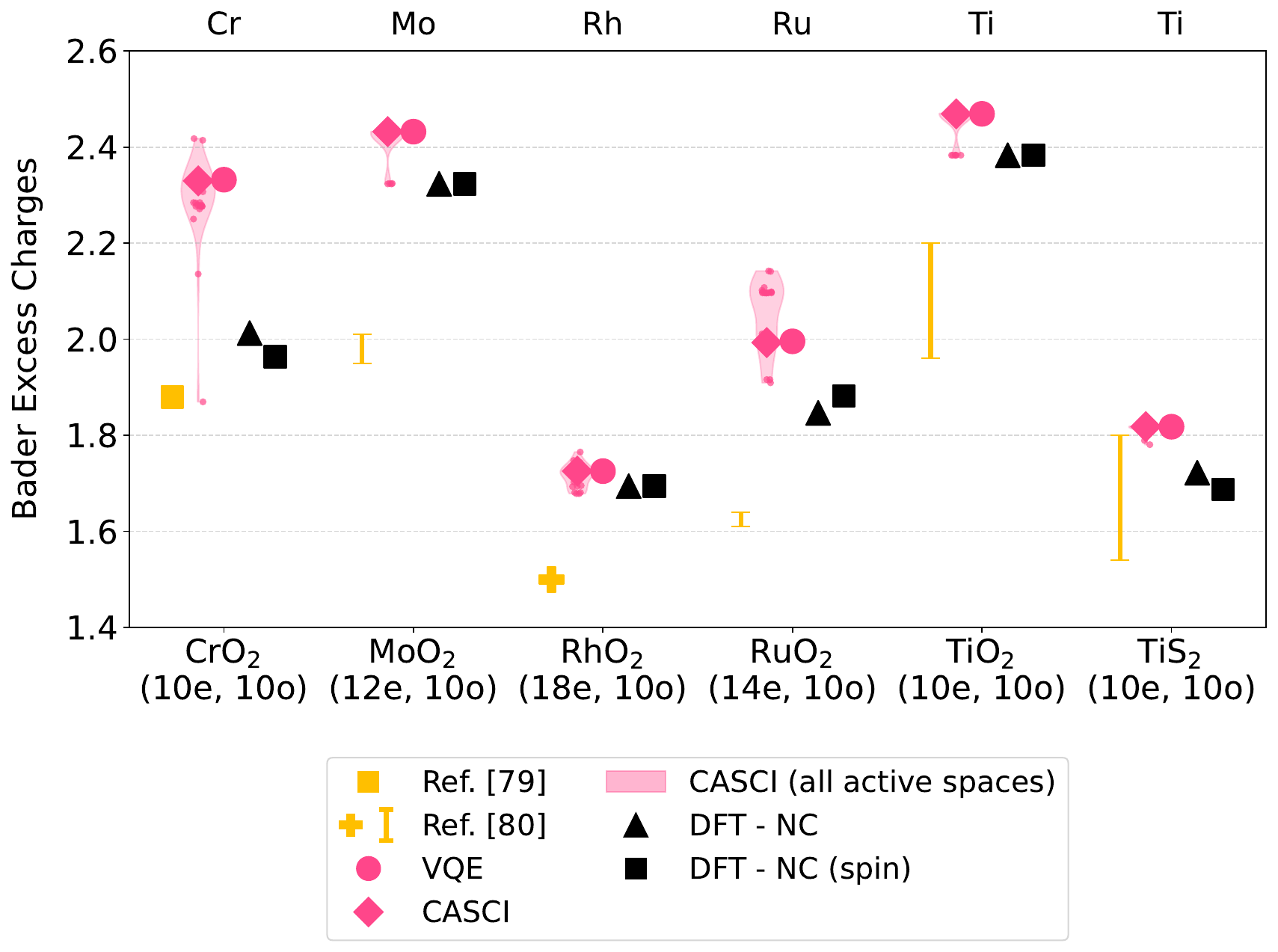}
    \caption{BECs on the different atom types in transition metal oxides (TMOs) and transition metal disulfide (TMS) for a representative active space for CASCI (pink diamonds) and VQE calculations (pink circles), and for various other active spaces using CASCI (pink shaded areas and pink dots). The BECs are also computed with DFT-NC. They are compared to Refs.~\cite{Lu2018Unraveling, Choudhuri2020Calculating}. The yellow square is the reference value of Ref.~\cite{Lu2018Unraveling}, and the yellow plus and bars are the reference data of Ref.~\cite{Choudhuri2020Calculating}, black triangles are DFT-NC results, and black squares are DFT-NC results from a spin-polarized calculation. The pink shaded areas show the distribution of the CASCI BECs over the computed active spaces. The chosen representative active spaces are given on the x-axis.}
    \label{fig:bader_TMO}
\end{figure}

The second set of systems -- \ce{CrO_2}, \ce{MoO_2}, \ce{RhO_2}, \ce{RuO_2}, \ce{TiO_2}, \ce{TiS_2} -- was selected specifically to study correlation effects DFT is not designed to capture. Reference BECs were obtained either from DFT+U calculations with a specific choice of U (\ce{CrO_2}, see Ref.~\cite{Lu2018Unraveling}) or from a scan over multiple values of U from Ref.~\cite{Choudhuri2020Calculating}, yielding the intervals reported in Figure~\ref{fig:bader_TMO}.
The scan in the Supporting Information of Ref.~\cite{Choudhuri2020Calculating} covers values of U from $0$ to $10\,\mathrm{eV}$, where $U=0$ corresponds to a standard DFT calculation without a Hubbard correction and to the lower end of each interval. With increasing U, the BEC increases for \ce{TiO_2}, \ce{TiS_2}, and \ce{MoO_2}, while it stays nearly constant for \ce{RuO_2} and \ce{RhO_2}~\cite{Choudhuri2020Calculating}.
In addition to spin-unpolarized DFT-NC calculations, we performed spin-polarized DFT-NC calculations for all six systems, with various starting magnetizations and show the lowest-energy solution among those tested as a black square in Fig.~\ref{fig:bader_TMO}.
These solutions have total (absolute) magnetic moments of $6.00$ ($6.19$), $0.00$ ($0.00$), $0.00$ ($0.00$), $1.61$ ($2.07$), $0.00$ ($0.00$), and $0.00$ ($1.05$)\,$\mu_\mathrm{B}$ per computational cell for \ce{CrO_2}, \ce{MoO_2}, \ce{RhO_2}, \ce{RuO_2}, \ce{TiO_2}, and \ce{TiS_2}, respectively. Thus, the lowest-energy solutions are ferromagnetic for \ce{CrO_2} and \ce{RuO_2} and nonmagnetic for \ce{MoO_2}, \ce{RhO_2}, and \ce{TiO_2}, while for \ce{TiS_2} the two Ti moments are parallel and compensated by antiparallel moments on the S atoms, so that the total magnetization vanishes but the absolute magnetization does not. We note that the total magnetization $M_\mathrm{tot}$ is related to the spin-projection expectation value $\braket{\hat S_z}$ of the computational cell via $M_\mathrm{tot} = 2\mu_\mathrm{B} \braket{\hat S_z}$ while no direct connection exists between the absolute magnetization and the quantum number $S$.
The total DFT magnetization should not be identified with the $S_z$ value used in a CASCI calculation. The former characterizes the independently optimized DFT solution, whereas the latter specifies the spin projection of the active-space wave function. 
Allowing spin polarization in the DFT-NC calculations has only a small effect on the BECs of the nonmagnetic \ce{MoO_2}, \ce{RhO_2}, \ce{TiO_2} systems, whereas more pronounced changes are observed for \ce{CrO_2}, \ce{RuO2} and \ce{TiS_2}. Thus, the influence of spin polarization on the BEC, is strongly system dependent but small and follows the magnetization we observe in our spin-unpolarized DFT-NC calculations.

The central question is therefore how the size of the active space affects the BECs and how the BEC changes obtained with CASCI and VQE differ from those obtained with DFT+U. Already at the DFT level, our spin-unpolarized DFT-NC BECs lie $0.18$ to $0.42\,\mathrm{e}$ above the $U=0$ BEC values of Ref.~\cite{Choudhuri2020Calculating}. We therefore compare the BEC changes of each method relative to its own DFT result rather than the absolute BECs.

\noindent The CASCI BECs of all computed active spaces are also shown for each system in Fig.~\ref{fig:as_scan} in Appendix~\ref{app:active_space_scan}. For all six systems, the BECs converge with the size of the active space, although the BECs of \ce{RhO_2} and \ce{RuO_2} alternate between two groups of values for the largest active spaces. In all representative active spaces, the VQE energies lie within $1.2\,\mathrm{mHa}$ of the CASCI energies, and the VQE BECs agree with the CASCI BECs within $0.003\,\mathrm{e}$.

For \ce{CrO_2}, the spin-unpolarized DFT-NC BEC on Cr is $2.01\,\mathrm{e}$ and the spin-polarized DFT-NC BEC is $1.96\,\mathrm{e}$, compared to the reference value of $1.88\,\mathrm{e}$. In the representative (10e, 10o) active space, VQE and CASCI both yield a BEC of $2.33\,\mathrm{e}$. Over all scanned active spaces, the CASCI BECs range from $1.87$ to $2.42\,\mathrm{e}$, and the active spaces with $11$ to $13$ spatial orbitals yield $2.27$ to $2.34\,\mathrm{e}$. Thus, the many-body treatment increases the BEC by about $0.3\,\mathrm{e}$ relative to spin-unpolarized DFT-NC, whereas spin polarization at the DFT-NC level decreases it by $0.05\,\mathrm{e}$. Since Ref.~\cite{Lu2018Unraveling} reports the BEC for \ce{CrO_2} only for a single value of U and not for $U=0$, the change caused by the Hubbard correction relative to DFT cannot be determined from Ref.~\cite{Lu2018Unraveling}, and we cannot compare it to the change caused by the many-body treatment.
The CASCI wave function at $S_z=0$ is dominated by six determinants of identical weight ($10.45\%$ each) (see Table~\ref{tab:casci_amplitudes}). An additional $S_z=S$ calculation shows that this multi-determinant character partly represents the spin configuration, while a noticeable multi-reference character remains at $S_z=S$.
The many-body treatment recovers this high-spin ground state within the active space from spin-unpolarized Kohn-Sham orbitals.

For \ce{RuO_2}, the spin-unpolarized and spin-polarized DFT-NC BECs are $1.85\,\mathrm{e}$ and $1.88\,\mathrm{e}$, respectively. In the representative (14e, 10o) active space, VQE yields a BEC of $2.00\,\mathrm{e}$ and CASCI a BEC of $1.99\,\mathrm{e}$. Over all computed active spaces, the CASCI BECs range from $1.91$ to $2.14\,\mathrm{e}$, and the active spaces with $11$ to $13$ spatial orbitals yield values around $2.00\,\mathrm{e}$ and around $2.10\,\mathrm{e}$. Thus, the many-body treatment increases the BEC by $0.15\,\mathrm{e}$ relative to spin-unpolarized DFT-NC in the representative active space, whereas the reference BEC changes only from $1.61$ to $1.64\,\mathrm{e}$ between $U=0$ and $U=10\,\mathrm{eV}$. The leading CASCI determinant has a weight of $62.82\%$ and the next determinant a weight of $14.00\%$, which is the most pronounced multi-reference character among the singlet ground states considered here.

For \ce{RhO_2}, the DFT-NC BEC of $1.69\,\mathrm{e}$ is unchanged by spin polarization, compared to the reference value of $1.50\,\mathrm{e}$, which is nearly independent of U. In the representative (18e, 10o) active space, VQE and CASCI both yield a BEC of $1.73\,\mathrm{e}$. Over all scanned active spaces, the CASCI BECs range from $1.68$ to $1.77\,\mathrm{e}$, and the active spaces with $11$ to $13$ spatial orbitals yield BECs from $1.70$ to $1.75\,\mathrm{e}$. The increase relative to DFT-NC, $0.03\,\mathrm{e}$ in the representative active space, is the smallest among the six systems, although the CASCI ground state contains, besides the leading determinant with $88.49\%$, two single excitations with $4.88\%$ each.

For \ce{TiS_2}, the spin-unpolarized and spin-polarized DFT-NC BECs of $1.72\,\mathrm{e}$ and $1.69\,\mathrm{e}$ lie within the reference interval of $1.54$--$1.80\,\mathrm{e}$. In the representative (10e, 10o) active space, VQE and CASCI both yield a BEC of $1.82\,\mathrm{e}$, and the CASCI BECs of all computed active spaces range from $1.78$ to $1.82\,\mathrm{e}$. The CASCI ground state is essentially single-reference with a weight of $99.93\%$ on the domninant determinant $\ket{\Psi_0}$. The increase of the BEC by $0.10\,\mathrm{e}$ relative to DFT-NC therefore mainly reflects the integer occupations of $\ket{\Psi_0}$ compared to the fractional occupations of the DFT-NC calculation. In Ref.~\cite{Choudhuri2020Calculating}, the BEC of \ce{TiS_2} increases from $1.54$ to $1.80\,\mathrm{e}$ between $U=0$ and $U=10\,\mathrm{eV}$, which is the largest change among the systems considered here.

For \ce{MoO_2} and \ce{TiO_2}, the DFT-NC BECs of $2.32\,\mathrm{e}$ and $2.38\,\mathrm{e}$ are unchanged by spin polarization and lie above the reference intervals of $1.95$--$2.01\,\mathrm{e}$ and $1.96$--$2.20\,\mathrm{e}$. In the representative (12e, 10o) and (10e, 10o) active spaces, VQE and CASCI yield BECs of $2.43\,\mathrm{e}$ and $2.47\,\mathrm{e}$, respectively. The computed active spaces yield two groups of BECs for both systems. A few active spaces reproduce the DFT-NC BEC within $10^{-3}\,\mathrm{e}$, and their CASCI ground states consist almost entirely of a single determinant, with weights above $98.8\%$ for \ce{MoO_2} and above $99.99\%$ for \ce{TiO_2}. All other active spaces, including all active spaces with $11$ to $13$ spatial orbitals, yield $2.43\,\mathrm{e}$ for \ce{MoO_2} and $2.47\,\mathrm{e}$ for \ce{TiO_2} within $0.01\,\mathrm{e}$. In the representative active spaces, the weight of the dominant determinant is $86.65\%$ for \ce{MoO_2} and $91.79\%$ for \ce{TiO_2}. Thus, a small admixture of excited determinants already increases the BECs by $0.11\,\mathrm{e}$ and $0.09\,\mathrm{e}$, respectively. In Ref.~\cite{Choudhuri2020Calculating}, the BECs increase by $0.06\,\mathrm{e}$ for \ce{MoO_2} and by $0.24\,\mathrm{e}$ for \ce{TiO_2} between $U=0$ and $U=10\,\mathrm{eV}$.

We note that a single reference in our CASCI and VQE calculations does not in general imply equality with the DFT density since the single reference uses integer occupations whereas the DFT calculation can use fractional occupations. For \ce{TiO_2}, the occupations of the spin-unpolarized DFT-NC calculation at the $\brm\Gamma$-point deviate from $0$ or $2$ by at most about $8.25\times10^{-9}$ electrons, and the CASCI and DFT-NC BECs agree within $10^{-4}$ electrons for the active spaces whose CASCI ground state consists almost entirely of the dominant determinant. For \ce{TiS_2}, \ce{CrO_2}, and \ce{RuO_2}, the DFT-NC calculations have fractional occupations at the $\brm\Gamma$-point.
In the representative active spaces, the CASCI ground states of \ce{MoO_2}, \ce{TiO_2}, \ce{RhO_2}, and \ce{TiS_2} are dominated by a single determinant with weights between $86.65\%$ and $99.93\%$. This dominance of a single determinant does not imply the absence of correlated electronic behavior in these compounds, which is well established in the literature for \ce{MoO_2}, \ce{TiO_2}, and \ce{TiS_2} \cite{Shevlin_Woodley_2010, Gulino_Parker_Jones_Egdell_1996, Scanlon_Watson_Payne_Atkinson_Egdell_Law_2010, Paxton_Thiên-Nga_1998, Werth2017Density}. Active spaces with more than $13$ spatial orbitals may be required to determine whether additional correlation effects appreciably affect their charge distributions, but such calculations are currently computationally intractable.

\begin{table*}[ht]
    \centering
    \caption{CASCI ground-state wave function contributions for the TMOs in the representative active spaces. We list only determinants with an absolute coefficient $|c_i| > 10^{-2}$, and among those only the leading determinants. $n$ is the number of determinants with $|c_i| > 10^{-2}$ and $\sum |c_i|^2$ is their summed weight. The orbitals are enumerated as follows: The first index is $1$ and each index refers to a spin-orbital where the first $M$ orbitals are spin-up, and the next $M$ orbitals are spin-down, where $M$ is the number of spatial orbitals in the active space. The total spin quantum number $S$ is also given for each system. $\ket{\Psi_0}$ is the single reference determinant used in VQE~(cf.~Sec.~\ref{sec:vqe}) which uses integer occupations and $\ket{\Psi^\prime_0}$ is the leading reference determinant of CrO$_2$ when setting $S_z=S$, with spin-orbitals $\{1,2,3,4,6,7,8,11,12,14\}$ occupied.}
    \label{tab:casci_amplitudes}
    
    \begin{tabular*}{\textwidth}{@{\extracolsep{\fill}} l r r l r r }
    \hline
    System & Determinant $|\psi_i\rangle$ & $|c_i|^2$ & System & Determinant $|\psi_i\rangle$ & $|c_i|^2$  \\
    \hline
     & & & & & \\
    \multicolumn{3}{l}{\textbf{TiS$_2$} (10e, 10o) \quad ($n=2$, $\sum |c_i|^2 = 99.95\%$, $S=0$)} & \multicolumn{3}{l}{\textbf{MoO$_2$} (12e, 10o) \quad ($n=23$, $\sum |c_i|^2 = 99.86\%$, $S=0$)} \\[2pt]
    
     $S_z=0$: & $\ket{\Psi_0}$ & $99.93\%$ &  & $\ket{\Psi_0}$ & $86.65\%$ \\
      & $\hat{a}_{5}\hat{a}_{15}\hat{a}^{\dagger}_{10}\hat{a}^{\dagger}_{20}\ket{\Psi_0}$ & $0.02\%$ &  & $\hat{a}_{5}\hat{a}^{\dagger}_{7}\ket{\Psi_0}$ & $5.31\%$ \\
      &  &  &  & $\hat{a}_{15}\hat{a}^{\dagger}_{17}\ket{\Psi_0}$ & $5.31\%$ \\
      &  &  &  & $\hat{a}_{6}\hat{a}_{16}\hat{a}^{\dagger}_{10}\hat{a}^{\dagger}_{20}\ket{\Psi_0}$ & $0.60\%$ \\
      &  &  &  & $\hat{a}_{4}\hat{a}^{\dagger}_{10}\ket{\Psi_0}$ & $0.34\%$ \\
     & & & & & \\[4pt]
    
    \multicolumn{3}{l}{\textbf{CrO$_2$} (10e, 10o) \quad ($n=68$, $\sum |c_i|^2 = 99.60\%$, $S=2$)} & \multicolumn{3}{l}{\textbf{RuO$_2$} (14e, 10o) \quad ($n=60$, $\sum |c_i|^2 = 99.58\%$, $S=0$)} \\[2pt]
    
     $S_z=0$: & $\hat{a}_{3}\hat{a}_{5}\hat{a}_{15}\hat{a}^{\dagger}_{7}\hat{a}^{\dagger}_{8}\hat{a}^{\dagger}_{16}\ket{\Psi_0}$ & $10.45\%$ &  & $\ket{\Psi_0}$ & $62.82\%$ \\
      & $\hat{a}_{3}\hat{a}_{5}\hat{a}_{15}\hat{a}^{\dagger}_{6}\hat{a}^{\dagger}_{7}\hat{a}^{\dagger}_{18}\ket{\Psi_0}$ & $10.45\%$ &  & $\hat{a}_{1}\hat{a}_{11}\hat{a}^{\dagger}_{8}\hat{a}^{\dagger}_{18}\ket{\Psi_0}$ & $14.00\%$ \\
      & $\hat{a}_{3}\hat{a}_{5}\hat{a}_{15}\hat{a}^{\dagger}_{6}\hat{a}^{\dagger}_{8}\hat{a}^{\dagger}_{17}\ket{\Psi_0}$ & $10.45\%$ &  & $\hat{a}_{4}\hat{a}_{14}\hat{a}^{\dagger}_{10}\hat{a}^{\dagger}_{20}\ket{\Psi_0}$ & $5.23\%$ \\
      & $\hat{a}_{5}\hat{a}_{13}\hat{a}_{15}\hat{a}^{\dagger}_{7}\hat{a}^{\dagger}_{16}\hat{a}^{\dagger}_{18}\ket{\Psi_0}$ & $10.45\%$ &  & $\hat{a}_{1}\hat{a}_{4}\hat{a}_{11}\hat{a}_{14}\hat{a}^{\dagger}_{8}\hat{a}^{\dagger}_{10}\hat{a}^{\dagger}_{18}\hat{a}^{\dagger}_{20}\ket{\Psi_0}$ & $2.60\%$ \\
      & $\hat{a}_{5}\hat{a}_{13}\hat{a}_{15}\hat{a}^{\dagger}_{8}\hat{a}^{\dagger}_{16}\hat{a}^{\dagger}_{17}\ket{\Psi_0}$ & $10.45\%$ &  & $\hat{a}_{4}\hat{a}_{14}\hat{a}^{\dagger}_{8}\hat{a}^{\dagger}_{18}\ket{\Psi_0}$ & $2.31\%$ \\
      & $\hat{a}_{5}\hat{a}_{13}\hat{a}_{15}\hat{a}^{\dagger}_{6}\hat{a}^{\dagger}_{17}\hat{a}^{\dagger}_{18}\ket{\Psi_0}$ & $10.45\%$ &  &  &  \\
     $S_z=S$: & $\ket{\Psi^\prime_0}$ & $62.70\%$ &  &  &  \\
      & $\hat{a}_{8}\hat{a}^{\dagger}_{9}\ket{\Psi^\prime_0}$ & $21.06\%$ &  &  &  \\
      & $\hat{a}_{14}\hat{a}^{\dagger}_{15}\ket{\Psi^\prime_0}$ & $6.52\%$ &  &  &  \\
      & $\hat{a}_{4}\hat{a}^{\dagger}_{5}\ket{\Psi^\prime_0}$ & $3.64\%$ &  &  &  \\
      & $\hat{a}_{8}\hat{a}_{14}\hat{a}^{\dagger}_{9}\hat{a}^{\dagger}_{15}\ket{\Psi^\prime_0}$ & $2.41\%$ &  &  &  \\
      & $\hat{a}_{4}\hat{a}_{8}\hat{a}^{\dagger}_{5}\hat{a}^{\dagger}_{9}\ket{\Psi^\prime_0}$ & $2.01\%$ &  &  &  \\
     & & & & & \\[4pt]
    
    \multicolumn{3}{l}{\textbf{RhO$_2$} (18e, 10o) \quad ($n=11$, $\sum |c_i|^2 = 99.99\%$, $S=0$)} & \multicolumn{3}{l}{\textbf{TiO$_2$} (10e, 10o) \quad ($n=6$, $\sum |c_i|^2 = 99.95\%$, $S=0$)} \\[2pt]
    
     $S_z=0$: & $\ket{\Psi_0}$ & $88.49\%$ &  & $\ket{\Psi_0}$ & $91.79\%$ \\
      & $\hat{a}_{6}\hat{a}^{\dagger}_{10}\ket{\Psi_0}$ & $4.88\%$ &  & $\hat{a}_{5}\hat{a}^{\dagger}_{7}\ket{\Psi_0}$ & $4.00\%$ \\
      & $\hat{a}_{16}\hat{a}^{\dagger}_{20}\ket{\Psi_0}$ & $4.88\%$ &  & $\hat{a}_{15}\hat{a}^{\dagger}_{17}\ket{\Psi_0}$ & $4.00\%$ \\
      & $\hat{a}_{1}\hat{a}_{11}\hat{a}^{\dagger}_{10}\hat{a}^{\dagger}_{20}\ket{\Psi_0}$ & $1.23\%$ &  & $\hat{a}_{5}\hat{a}_{15}\hat{a}^{\dagger}_{7}\hat{a}^{\dagger}_{17}\ket{\Psi_0}$ & $0.13\%$ \\
     & & & & & \\[4pt]
    
    \hline
    \end{tabular*}
\end{table*}

The total spin quantum numbers $S$ reported in Table~\ref{tab:casci_amplitudes} refer to the active-space CASCI wave function and reveal that the system with a marginal reference determinant weight, CrO$_2$, is a high-spin state, namely a quintet ($S=2$). All other systems are singlets ($S=0$) with reference determinant weights between $62.82\%$ (RuO$_2$) and $99.93\%$ (TiS$_2$). We note that the values for $S$ are the same for both VQE and CASCI results for all TMO systems. In a high-spin state, a single spin-restricted reference determinant has a poor overlap with the true ground state within the active space \cite{Erakovic2025High}.
To verify this, we performed an additional CASCI calculation for \ce{CrO_2} with the spin projection fixed to $S_z=S$, using the same active space (see Table~\ref{tab:casci_amplitudes}) and resulting in the same energy as with $S_z=0$. The leading determinant then carries $62.70\%$ of the wave function norm, and the next two determinants carry $21.06\%$ and $6.52\%$. Thus, the $S_z=0$ multi-determinant character of \ce{CrO_2} partly represents the spin configuration, while a multi-reference character remains at $S_z=S$. We note that setting $S_z=S$ does not change the BEC.
In the representative active spaces, VQE and CASCI increase the BECs relative to spin-unpolarized DFT-NC for all six systems, by $0.32\,\mathrm{e}$ (\ce{CrO_2}), $0.15\,\mathrm{e}$ (\ce{RuO_2}), $0.11\,\mathrm{e}$ (\ce{MoO_2}), $0.10\,\mathrm{e}$ (\ce{TiS_2}), $0.09\,\mathrm{e}$ (\ce{TiO_2}), and $0.03\,\mathrm{e}$ (\ce{RhO_2}). Among the singlet systems, the largest change in BEC (\ce{RuO_2}) coincides with the lowest reference determinant weight. But we cannot observe a general connection between change in BEC and detereminant weights. For example, \ce{RhO_2} has a lower reference determinant weight than \ce{TiO_2} and \ce{TiS_2} but the smallest change. Additionally, for \ce{TiS_2} the change mainly reflects the integer occupations of the dominant determinant compared to fractional occupations in the DFT-NC calculations. For \ce{CrO_2}, the determinant multiplicity in the $S_z=0$ sector partly represents the high-spin configuration. Consequently, the BEC changes cannot be attributed solely to multi-reference correlation on the basis of the $S_z=0$ determinant weights.
These CASCI results should not be identified directly with the spin-polarized DFT-NC calculations shown in Fig.~\ref{fig:bader_TMO}. Spin-polarized DFT allows for spin-symmetry breaking which may change both the spin density and the spin-summed charge density. By contrast, the $S_z=0$ and $S_z=S$ CASCI calculations correspond to different spin projections of the same total-spin multiplet and yield identical spin-summed densities and BECs. Consequently, neither the number of important determinants in the $S_z=0$ configuration nor the CASCI value of $S$ directly predicts the change in BEC produced by spin-polarized DFT.

\section{Discussion}
We present the computation of BECs using our many-body post-processing framework, which builds and solves a many-body Hamiltonian from the results of a DFT calculation. We calculate BECs for various systems with various computational methods (DFT, CASCI and VQE) and compare the results to literature values~\cite{PaskasMamula2014Electronic, Lu2018Unraveling, Choudhuri2020Calculating}. The reference calculations we compare to are done using DFT ~\cite{PaskasMamula2014Electronic} or DFT+U ~\cite{Lu2018Unraveling, Choudhuri2020Calculating} using PAW pseudopotentials and k-meshes, while all our results are based on norm-conserving pseudopotentials and $\brm\Gamma$-point calculations. All reference DFT results were obtained with spin-unpolarized calculations. Only the reference values for CrO$_2$, taken from \cite{Lu2018Unraveling}, have been obtained using a spin-polarized calculation.

For the \ce{MgH_2} systems,  the many-body post-processing has no significant impact on the BEC of the central dopant atom. These systems have been extensively studied with standard DFT in the literature and we considered them as being weakly correlated \cite{PaskasMamula2014Electronic, Lyu2023OnTheCatalytic, Chen2004Alloying, Zhang2017Enhanced}, and accordingly no improvement over the single-reference descriptions is expected. The agreement between DFT, CASCI and VQE charges and the reference values therefore serves primarily as a validation of the present computational framework, confirming that the active-space construction and the many-body treatment are internally consistent. The exception is \ce{MgH_2}:\ce{Fe}, where the BEC is sensitive to the active space size: with the (6e, 9o) active space, CASCI and VQE underestimate the reference, whereas expanding to (18e, 13o) brings the CASCI charge into closer agreement with the reference. We can therefore assess our computational framework as being accurate in the calculation of BECs.

Our results on the BECs of \ce{TMO_2} show significant differences between the compared methods.
Already at the DFT level, our DFT-NC BECs lie $0.18$ to $0.42\,\mathrm{e}$ above the values for $U=0$ of Ref.~\cite{Choudhuri2020Calculating}. We attribute these changes to the differences in the used pseudopotentials and the k-point sampling. We therefore compare the BEC changes of each method relative to its own DFT result. With increasing size of the active space, the CASCI BECs converge for all six systems, although the BECs of \ce{RhO_2} and \ce{RuO_2} alternate between two groups of values for the largest active spaces. In the representative active spaces, VQE and CASCI increase the BECs relative to spin-unpolarized DFT-NC for all six systems, by $0.03$ to $0.32\,\mathrm{e}$, and the VQE BECs agree with the CASCI BECs within $0.003\,\mathrm{e}$. DFT+U also increases the BECs with increasing U for \ce{TiO_2}, \ce{TiS_2}, and \ce{MoO_2}, by up to $0.24$, $0.26$, and $0.06\,\mathrm{e}$ at $U=10\,\mathrm{eV}$, while the BECs of \ce{RuO_2} and \ce{RhO_2} are nearly independent of U~\cite{Choudhuri2020Calculating}. Thus, our many-body approach and DFT+U shift the BECs in the same direction, but the size of the shifts depends on the system. Our many-body approach yields its largest shifts for \ce{CrO_2} and \ce{RuO_2}, while DFT+U yields its largest shifts for \ce{TiS_2} and \ce{TiO_2} and leaves the BEC of \ce{RuO_2} nearly unchanged. Since the frozen-core electrons are described at a HF-like mean-field level (see Sec.~\ref{sec:frozen_core_approx}), part of the BEC increases may reflect the tendency of mean-field descriptions with exact exchange to over-localize electrons \cite{MoriSanchez2008Localization, Li2017Localized}, as discussed in Sec.~\ref{sec:limitations}. The common direction of the BEC changes obtained with our approach and with DFT+U therefore does not imply that both capture the same physical effect. The two approaches also treat electronic correlation differently. DFT+U adds an on-site correction at a mean-field level whose strength is set by U, while our approach treats the correlation of the electrons in the active space explicitly and describes the remaining electrons at a mean-field level.

The BEC changes do not exhibit a one-to-one correspondence with the multi-reference character of the CASCI ground states. Among the singlet systems, \ce{RuO_2} shows both the largest BEC change and the lowest reference determinant weight, whereas the BEC change of \ce{TiS_2} arises mainly from the integer occupations of the reference determinant. For \ce{MoO_2} and \ce{TiO_2}, already a small admixture of excited determinants increases the BEC by about $0.1\,\mathrm{e}$. For \ce{CrO_2}, the determinant multiplicity in the $S_z=0$ sector partly represents the high-spin state.

In our calculations, allowing spin polarization at the DFT-NC level decreases the Cr BEC from $2.01\,\mathrm{e}$ to $1.96\,\mathrm{e}$, while CASCI and VQE based on the spin-unpolarized KS orbitals yield $2.33\,\mathrm{e}$. The CASCI solution has $S=2$, showing that the active-space calculation can recover a high-spin state from spin-unpolarized KS orbitals. Thus, spin polarization at the DFT-NC level and the many-body treatment change the Cr BEC in opposite directions. The difference between spin-polarized DFT-NC and the many-body results cannot be assigned uniquely to electronic correlation, because the calculations also differ in their many-body treatment and frozen-core approximation.

Overall, the magnitude of the BEC change caused by the many-body treatment is system dependent and cannot be inferred directly from either the CASCI total spin or the number of important determinants in the $S_z=0$ representation. The differences of the absolute BECs with respect to the references may additionally reflect the use of norm-conserving rather than PAW pseudopotentials, $\brm{\Gamma}$-point sampling rather than k-meshes, and the relaxation of the atomic positions within our DFT-NC treatment.

By comparing the BEC changes from our many-body calculations to those of reference DFT+U data, we evaluate how a correlated active-space treatment in combination with VQE changes the DFT description of the charge distribution for strongly correlated systems. As a general conclusion, the many-body treatment within our framework at the $\brm{\Gamma}$-point increases the BECs relative to DFT for all six systems. For several systems, this is the same direction as the change caused by the Hubbard correction in DFT+U, but the size of the increase depends on the system and the common direction does not imply a common physical origin. Our approach differs from DFT+U in several practical aspects, although it requires a computationally expensive many-body calculation on top of DFT. First, DFT+U calculations are generally time-consuming and cumbersome because the Hubbard U parameter must be tuned carefully to either fit experimental data or using computationally costly methods such as linear-response~\cite{Cococcioni2005Linear}. Further, the U value is not easily transferable between systems since it depends heavily on local stoichiometry and structure~\cite{Cai2024Predicting}. One must also carefully determine on which orbitals the Hubbard correction should be applied, since $s$ and $p$ orbitals can sometimes also contribute to the strongly correlated nature of a system~\cite{Macke2024OrbitalResolved}.

In contrast, our framework is straightforward: we perform a spin-unpolarized DFT calculation at the $\brm{\Gamma}$-point and apply our post-processing code within a selected active space to solve the many-body problem. No parameters in the Hamiltonian need to be tuned. However, the BECs depend on the choice of the active space and on the frozen core approximation.
Additionally, our method is naturally suited for use with quantum resources, when employing VQE on quantum hardware, whereas DFT is optimized for classical computers, although some works do propose the use of quantum computers for DFT calculations~\cite{Senjean2023Toward, Ko2023Implementation}. Furthermore, our workflow can readily be extended to take a spin-polarized k-mesh calculation as input, since we rely only on the Kohn-Sham orbitals and crystal structure information to build the many-body Hamiltonian. 

The total spin values obtained from our CASCI wave functions (Table~\ref{tab:casci_amplitudes}) indicate that CrO$_2$ exhibits a high-spin ground state within the active space, while all other TMO systems are singlets.
For CrO$_2$, an $S_z=S$ CASCI calculation yields an identical BEC and shows that the multi-reference character of the $S_z=0$ result is partly related to the spin configuration, while a multi-reference character remains at $S_z=S$.
We leave the investigation of DFT inputs based on different pseudopotentials and spin-polarized k-meshes as future work.
Additional future work involves introducing orbital optimization in terms of classical complete active space self consistent field~(CASSCF) and hybrid quantum algorithms \cite{Mizukami2020Orbital,Bierman2023Improving, Yalouz2022Analytical} to access the usefulness of utilizing KS orbitals which were optimized in a DFT simulation. Further, we need to use different ansätze and optimization strategies within VQE and other hybrid quantum algorithms like ADAPT-VQE \cite{Grimsley2019AnAdaptive} to be able to increase the active-space sizes for our simulations. Other embedding techniques like the WF-in-DFT embedding \cite{Manby2012ASimple, Lee2019ProjectionBased} should also be addressed in the future to more accurately describe the electronic structure and better utilize the prior DFT calculations that act as input to our framework.

\section*{Code availability}
    The source code of Dopyqo 
    is available on GitHub: \url{https://github.com/dlr-wf/Dopyqo}. This includes the calculation of many-body Hamiltonians from Quantum ESPRESSO output files, ground state calculations with classical algorithms and VQE, and computing Bader charges with the Bader charge analysis code \cite{Arnaldsson2023BaderCode}. Some minimal examples are provided to make the experiments conducted in this work reproducible.

\section*{Acknowledgments}
    We thank Daniel Barragan-Yani, Max Haas, and Thierry Kaldenbach for helpful comments on the manuscript.
    
\section*{Funding}
    This project was made possible by the DLR Quantum Computing Initiative and the Federal Ministry for Research, Technology and Space; \url{https://qci.dlr.de/quanticom}.

\section*{Author contributions}
    E.S. conceived the computational framework Dopyqo and implemented the calculation of matrix elements from Quantum ESPRESSO output files, the generation of Hamiltonians, and how the Hamiltonian is solved via VQE, CASCI.
    A.R. implemented the calculation of Bader charges using ground states and the Bader charge analysis code \cite{Arnaldsson2023BaderCode}.
    G.B. selected the Bader charges as the property to compute, selected the material systems, and performed the DFT and Bader charge calculations for the \ce{MgH2} systems.
    E.S. performed the DFT and Bader charge calculations for the TMO systems.
    E.S. and G.B. discussed, analysed and interpreted the results, and wrote the manuscript.

\section*{Competing interests}
    A patent application filed by the German Aerospace Center (Deutsches Zentrum für Luft- und Raumfahrt e.V., DLR), currently pending with the German Patent and Trade Mark Office (Deutsches Patent- und Markenamt, DPMA), covers aspects of this work. The authors declare no other financial or non-financial competing interests.

\printbibliography

\onecolumn

\clearpage
\appendix

\part*{Appendix}
\section{Symmetries of matrix elements} \label{app:symmeries-matrix-elements}
    Here we provide the symmetries of the matrix elements $h_{tu}$ and $h_{tuvw}$ which are independent of the $\brm{k}$-point \cite{Clinton2024Towards}. For complex wavefunctions we find
    \begin{equation} \label{app:eq:symm-mat-elems-complex}
        \begin{aligned}
            h_{pq}&=h^*_{qp} \quad&&\mathrm{(hermiticity)},\\
            h_{pqrs}&=h_{qpsr} \quad&&\mathrm{(swap\ symmetry)},\\
            h_{pqrs}&=h^*_{srqp} \quad&&\mathrm{(hermiticity)},\\
            h_{pqrs}&=h^*_{rspq} \quad&&\mathrm{(hermiticity+swap)}.
        \end{aligned}
    \end{equation}
    For real wavefunctions these simplify to
    \begin{equation} \label{app:eq:symm-mat-elems-real}
        \begin{aligned}
            h_{pq}&=h_{qp}\\
            h_{pqrs}&=h_{qpsr}=h_{srqp}=h_{rspq}\\
            &=h_{sqrp}=h_{prqs}=h_{rpsq}=h_{qspr}\,.
        \end{aligned}
    \end{equation}

\section{Local pseudopotential matrix element derivation} \label{app:local-pp-derivation}
    We assume the local pseudopotential $\hat{V}^{(I)}_\mathrm{loc}$ for the nucleus with index $I$, charge $Z_I$, and position $\brm R_I$ is given as a long-range radial-dependent real-space function  $V^{(I)}_\mathrm{loc}(r)=\bra{\brm{r}}\hat{V}^{(I)}_\mathrm{loc}\ket{\brm{r}}$, with its Fourier-transform being $\bra{\brm{G}_1}\hat{V}^{(I)}_\mathrm{loc}\ket{\brm{G}_2}$. We now derive Eqs.~\eqref{eq:v-loc-pw} and \eqref{eq:v-loc-pw-zero-mom-lim}. To handle the singularity at $\brm{G}_1=\brm{G}_2$, we use
    \begin{equation} \label{app:eq:vloc-plus-minus-erf}
        V^{(I)}_\mathrm{loc}(r) \to V^{(I)}_\mathrm{loc}(r) + \frac{Z_I\, \mathrm{erf}(r)}{r} - \frac{Z_I\, \mathrm{erf}(r)}{r} - \Phi^{(I)}_b(r)\,, 
    \end{equation}
    where, on the right hand side, the first two terms together are short-ranged and, therefore, non-singular for $\brm{G}_1=\brm{G}_2$. The electronic charges in the background, which are keeping the computational cell charge neutral, make the remaining two terms also non-singular. The potential $\Phi^{(I)}_b(r)$ is modelling this electronic background for the $I$-th nucleus and solves the Poisson equation for a constant charge density $\rho^{(I)}_b$:
    \begin{equation}
        \nabla^2 \Phi^{(I)}_b(r) = -\rho^{(I)}_b = +\frac{Z_I}{V}\,,
    \end{equation}
    and the Fourier transform of $\Phi^{(I)}_b(r)$ is
    \begin{equation}
        \tilde\Phi^{(I)}_b(\brm G) = -\frac{4\pi Z_I \delta(\brm G)}{VG^2}\,,
    \end{equation}
    where $\delta(\brm{G})$ is the Dirac delta distribution.
    Alternatively, in Eq.~\eqref{app:eq:vloc-plus-minus-erf}, one could also add and subtract $\frac{Z_I}{r}$, or any function that behaves like $\frac{Z_I}{r}$ for large $r$. We use the error function since $\mathrm{erf}(r)/r$ is non-singular for $r\to0$. 
    The error function $\mathrm{erf}(r)$ is defined as
    \begin{equation}
        \mathrm{erf}(r) = \frac{2}{\sqrt\pi} \int_0^r e^{-x^2} \mathrm{d}x\,.
    \end{equation}
    The Fourier transform of the error function is found to be
    \begin{equation}
        \int_\mathbb{R} \mathrm{erf}(r)\,e^{-iGr}\,\mathrm{d}r = -2i\,\frac{e^{-G^2/4}}{G}\,.
    \end{equation}
    From this, using Eq.~\eqref{eq:fourier-transform-of-periodic-function-our-case} and recognizing that $\mathrm{erf}(r)\sin(Gr)$ is an even function, it follows
    \begin{equation} \label{app:eq:fourier-trafo-of-erf-over-r}
        \frac1V \int_{\mathbb{R}^3}  \frac{\mathrm{erf}(r)}{r}\,e^{-i\brm{G}\cdot\brm{r}}\,\mathrm{d}\brm{r} = \frac{4\pi}{V} \int_0^\infty \mathrm{erf}(r)\,\frac{\sin(Gr)}{G}\,\mathrm{d}r=-\frac{4\pi}{V} \frac1{2G} \Im\left(\int_{-\infty}^{\infty} \mathrm{erf}(r)\,e^{-iGr}\,\mathrm{d}r\right)=\frac{4\pi}{V} \frac{e^{-G^2/4}}{G^2}\,,
    \end{equation}
    where $\Im(x)$ is the imaginary part of $x$. With this and using the spherical symmetry of $V^{(I)}_\mathrm{loc}(r)$ we find
    \begin{equation} \label{app:eq:v-loc-with-erf}
        \begin{aligned}
            \bra{\brm{G}_1}\hat{V}^{(I)}_\mathrm{loc}\ket{\brm{G}_2} &= \int_{\mathbb{R}^3} \left(V^{(I)}_\mathrm{loc}(r) + \frac{Z_I\, \mathrm{erf}(r)}{r}\right)\,e^{-i\left(\brm{G}_1-\brm{G}_2\right)\cdot\brm{r}}\, \mathrm{d}\brm{r} - \frac{4\pi Z_I}{V} \frac{e^{-\left(\brm{G}_1-\brm{G}_2\right)^2/4}}{\left(\brm{G}_1-\brm{G}_2\right)^2} - \tilde\Phi^{(I)}_b(\brm{G}_1-\brm{G}_2)\\
            &=\frac{4\pi}{V} \int_0^\infty r\,\left(V^{(I)}_\mathrm{loc}(r) + \frac{Z_I\, \mathrm{erf}(r)}{r}\right)\,\frac{\sin\left(\left|\brm{G}_1-\brm{G}_2\right|r\right)}{\left|\brm{G}_1-\brm{G}_2\right|} \mathrm{d}r - \frac{4\pi Z_I}{V} \frac{e^{-\left(\brm{G}_1-\brm{G}_2\right)^2/4}}{\left(\brm{G}_1-\brm{G}_2\right)^2} - \tilde\Phi^{(I)}_b(\brm{G}_1-\brm{G}_2)\,,
        \end{aligned}
    \end{equation}
    where we used $\bra{\brm{r}}\hat{V}^{(I)}_\mathrm{loc}\ket{\brm{r}^{\prime}}=V^{(I)}_\mathrm{loc}(r)\,\delta(\brm{r}-\brm{r}^{\prime})$ and $x=|\brm{x}|$. Equation~\eqref{app:eq:v-loc-with-erf} is non-singular for $\brm{G}_1=\brm{G}_2$ and can be used to calculate all matrix elements $\bra{\brm{G}_1}\hat{V}^{(I)}_\mathrm{loc}\ket{\brm{G}_2}$. The expression for $\brm{G}_1=\brm{G}_2$ of Eq.~\eqref{app:eq:v-loc-with-erf} is
    \begin{equation} \label{app:eq:v-loc-zero-mom-limit}
        \left.\bra{\brm{G}_1}\hat{V}^{(I)}_\mathrm{loc}\ket{\brm{G}_2}\right|_{\brm{G}_1=\brm{G}_2} = \frac{4\pi}{V} \int_0^\infty r^2\,\left(V^{(I)}_\mathrm{loc}(r) + \frac{Z_I}{r}\right) \mathrm{d}r\,,
    \end{equation}
    where we used Eq.~\eqref{app:eq:fourier-trafo-of-erf-over-r} and $\mathrm{erf}(r)=1-\mathrm{erfc}(r)$ with the complementary error function $\mathrm{erfc}(r)$. This result can also be obtained when using $\frac{Z_I}{r}$ instead of $\frac{Z_I\,\mathrm{erf}(r)}{r}$ in Eq.~\eqref{app:eq:vloc-plus-minus-erf}.
    In practice we use Eq.~\eqref{app:eq:v-loc-zero-mom-limit} to compute all matrix elements $\bra{\brm{G}_1}\hat{V}^{(I)}_\mathrm{loc}\ket{\brm{G}_2}$ where $\brm{G}_1=\brm{G}_2$ and use Eq.~\eqref{app:eq:v-loc-with-erf} to compute the remaining matrix elements. We note that Eqs.~\eqref{app:eq:v-loc-with-erf} and \eqref{app:eq:v-loc-zero-mom-limit} are only valid for $\brm R_I=\brm 0$, for $\brm R_I\neq\brm0$ we need to apply the Fourier shift theorem resulting in
    \begin{equation}
        \bra{\brm{G}_1}\hat{V}^{(I)}_\mathrm{loc}\ket{\brm{G}_2} \to e^{-i(\brm{G}_1-\brm{G}_2)\cdot \brm R_I} \bra{\brm{G}_1}\hat{V}^{(I)}_\mathrm{loc}\ket{\brm{G}_2}\,.
    \end{equation}

\section{Nuclear interaction derivation} \label{app:nuclear-int-ewald-derivation}
    We use the Ewald method \cite{Ewald1921DieBerechnung} to derive Eq.~\eqref{eq:nuclear-repulsion-energy-ewald}. For this we first define a charge distribution $\rho_{[I]}(r)$ for all $N$ nuclei in the computational cell except the $I$-th one,
    \begin{equation} \label{app:eq:nucl-charge-density-w-gaussians}
        \rho_{[I]}(\brm{r}) = \sum_{J=1}^N \sideset{}{'}\sum_{\brm{T}} Z_J \delta(\brm{r}-\brm{R}_J-\brm{T}) - \rho_b = \sum_{J=1}^N \sideset{}{'}\sum_{\brm{T}} \left[Z_J \delta(\brm{r}-\brm{R}_J-\brm{T}) - \rho^G_{J,\brm{T}}(\brm{r})\right] - \rho^G_{I,\brm{0}}(\brm{r}) + \sum_{J=1}^N \sum_{\brm{T}} \rho^G_{J,\brm{T}}(\brm{r}) + \rho_b\,,
    \end{equation}
    where we sum over all lattice translation vectors $\brm{T}$ and the prime in the summation indicates that we omit self-interaction terms with $\brm{T}=\bm{0}$ and $I=J$. $\rho_b$ is a constant density modelling the electronic background with
    \begin{equation}
        \rho_b = -\sum_I \frac{Z_I}{V}\,.
    \end{equation}
    We added and subtracted Gaussian charge distributions with
    \begin{equation}
        \rho^G_{I,\brm{T}}(\brm{r}) = \left(\frac{\sigma}{\pi}\right)^{3/2} Z_I\, e^{\sigma \left(\brm{r}-\brm{R}_I-\brm{T}\right)^2}\,,
    \end{equation}
    where $\sigma>0$. Now we employ the Ewald method to calculate the electrostatic potential generated by the first two terms in Eq.~\eqref{app:eq:nucl-charge-density-w-gaussians} in real space and evaluate the remaining two terms in reciprocal space. The result will be independent of the parameter $\sigma$, but in practice $\sigma$ can be chosen such that the real and reciprocal parts converge most rapidly.
    
    We begin with the real space part. To obtain the electrostatic potential $\phi^R_{[I]}(\brm{r})$ we solve the Poisson equation
    \begin{equation}
        \nabla^2 \phi^R_{[I]}(\brm{r}) = - \left\{\sum_{J=1}^N \sideset{}{'}\sum_{\brm{T}} \left[Z_J \delta(\brm{r}-\brm{R}_J-\brm{T}) - \rho^G_{J,\brm{T}}(\brm{r})\right] - \rho^G_{I,\brm{0}}(\brm{r})\right\}\,.
    \end{equation}
    The electrostatic potential $\phi^G_{I,\brm{T}}(\brm{r})$ for one Gaussian charge distribution $\rho^G_{I,\brm{T}}(\brm{r})$ is
    \begin{equation}
        \phi^G_{I,\brm{T}}(\brm{r}) = Z_I\,\frac{\mathrm{erf}\left(\sqrt\sigma \left|\brm{r}-\brm{R}_I-\brm{T}\right|\right)}{\left|\brm{r}-\brm{R}_I-\brm{T}\right|}\,,
    \end{equation}
    with the error function $\mathrm{erf}(x)$, while the electrostatic potential $\phi^\delta_{I,\brm{T}}(\brm{r})$ of a point charge $Z_I \delta(\brm{r}-\brm{R}_I-\brm{T})$ is given by
    \begin{equation}
        \phi^\delta_{I,\brm{T}}(\brm{r}) = \frac{Z_I}{\left|\brm{r}-\brm{R}_I-\brm{T}\right|}\,.
    \end{equation}
    With this we find
    \begin{equation}
        \phi^R_{[I]}(\brm{r}) = \sum_{J=1}^N \sideset{}{'}\sum_{\brm{T}} \frac{Z_J\,\mathrm{erfc}\left(\sqrt\sigma \left|\brm{r}-\brm{R}_J-\brm{T}\right|\right)}{\left|\brm{r}-\brm{R}_J-\brm{T}\right|} - \frac{Z_I\,\mathrm{erf}\left(\sqrt\sigma \left|\brm{r}-\brm{R}_I\right|\right)}{\left|\brm{r}-\brm{R}_I\right|}\,,
    \end{equation}
    where we used the complementary error function $\mathrm{erfc}(x)=1-\mathrm{erf}(x)$.
    
    To compute the reciprocal space part we solve the Poisson equation in reciprocal space:
    \begin{equation}
        \tilde{\phi}(\brm G) = \frac{4\pi\tilde\rho(\brm G)}{G^2}\,
    \end{equation}
    where $\tilde\phi(\brm G)$ and $\tilde\rho(\brm G)$ are the electrostatic potential and charge density in reciprocal space, respectively. We first Fourier transform $\sum_{\brm{T}}\rho^G_{I,\brm{T}}(\brm{r})$ using Eq.~\eqref{eq:fourier-transform-of-periodic-function-our-case} by integrating over the computational cell $C$:
    \begin{equation}
        \tilde{\rho}^G_{I}(\brm{G}) = \frac{1}{V} \int_{C} \sum_{\brm{T}} \rho^G_{I,\brm{T}}(\brm{r})\,e^{-i\brm{G} \cdot \brm{r}} \mathrm{d}\brm{r} = \frac{Z_I}{V}\,e^{-i\brm{G}\cdot\brm{R}_I - G^2/(4\sigma)}\,.
    \end{equation}
    With the Fourier transform of the electronic background $\rho_b$ being $\rho_b\delta(\brm{G})/V$ where $\delta(\brm{G})$ is the Dirac delta distribution, we obtain the electrostatic potential
    \begin{equation}
        \tilde\phi^G(\brm G) = \frac{4\pi}{V G^2} \left(\sum_I \tilde{\rho}^G_{I,\brm{T}}(\brm{G}) + \rho_b\delta(\brm{G}) \right)\,.
    \end{equation}
    Fourier transforming $\tilde\phi^G(\brm G)$ back into real space and carefully handling the $\brm G=\brm 0$ term, using the Taylor series of $e^{- G^2/(4\sigma)}$, yields $\phi^G(\brm r)$ with
    \begin{equation}
        \phi^G(\brm r) = \sum_I \frac{4\pi Z_I}{V}\sum_{\brm G\neq0}\frac{1}{G^2}\,e^{-i\brm G\cdot\brm R_I - G^2/(4\sigma)} e^{i\brm G\cdot\brm r} - \frac{\pi}{\sigma V} \sum_I Z_I\,.
    \end{equation}
    To now obtain the result from Eq.~\eqref{eq:nuclear-repulsion-energy-ewald}, we calculate the electrostatic energy $E_{\mathrm{n\text-n}}$ with
    \begin{equation}
        \begin{aligned}
            E_{\mathrm{n\text-n}} = \frac12 \sum_{I=1}^N Z_I \left(\phi^R_{[I]}(\brm R_I) + \phi^G(\brm R_I)\right) =& \frac{1}{2} \sum_{I,J,\bm{T}} Z_I Z_J \frac{\mathrm{erfc}(\sqrt{\sigma}\left|\brm{R_I} - \brm{R_J} - \brm{T}\right|)}{\left|\brm{R_I} - \brm{R_J} - \brm{T}\right|} - \sqrt{\frac{\sigma}{\pi}} \sum_I Z_I^2\\
            &+\frac{2\pi}{V} \sum_{\brm{G} \neq \brm{0}} \sum_{I,J} \frac{Z_I Z_J}{G^2} e^{i \brm{G} \cdot (\brm{R_I} - \brm{R_J}) - G^2/(4 \sigma)} - \frac{\pi}{2 \sigma V}  \left(\sum_I Z_I\right)^2\,,
        \end{aligned}
    \end{equation}
    where we used that $\lim_{x\to0}\,\mathrm{erf}(ax)/x = 2a/\sqrt\pi$.

\section{Spin-resolved frozen core approximation} \label{app:frozen-core-spin}
    In this section we derive the spin-polarized version of the frozen core energy and potential, which can be used to obtain Eq.~\eqref{eq:frozen-core-energy} and Eq.~\eqref{eq:frozen-core-potential}.
    In the following we use the indices $\{a,b,c,d\}$ for orbitals in the environment, $\{t,u,v,w\}$ for orbitals in the active space and $\{i,j,k,l\}$ for general orbitals. These indices run over spatial orbitals. We use $\sigma$ and $\tau$ as indices for the two different spin channels, i.e., $\sigma,\tau\in\{\uparrow,\downarrow\}$. We divide a state $\ket{\psi}$ into a environment ($\ket{\psi_\mathrm{env}}$) and active space ($\ket{\psi_\mathrm{active}}$) part where the environment is assumed to be fully occupied:
    \begin{equation}
        \ket{\psi}=\ket{\psi_\mathrm{env}}\otimes\ket{\psi_\mathrm{active}}\,.
    \end{equation}
    We explicitly consider spin in the operators and the matrix elements, i.e.,
    \begin{equation}
        \begin{gathered}
            a^{(\dagger)}_{i}\to a^{(\dagger)}_{i,\sigma}\,,\\
            h_{ij}\to h_{ij,\sigma}\,,\\
            h_{ijkl}\to h_{ijkl,\sigma\tau}=\int \int \psi^\ast_t(\brm{r}_1,\sigma)\,\psi^\ast_u(\brm{r}_2,\tau)\,V_\text{e-e}(\brm r_1, \brm r_2)\,\psi_v(\brm{r}_2,\tau)\,\psi_w(\brm{r}_1,\sigma) \mathrm{d}\brm{r}_1 \mathrm{d}\brm{r}_2\,,
        \end{gathered}
    \end{equation}
    with $V_\text{e-e}(\brm r_1, \brm r_2)$ defined in Eq.~\eqref{eq:periodic-e-e-interaction}. We use that the eigenvalues of $\hat{n}_{i, \sigma}=a^{\dagger}_{i, \sigma}a_{i, \sigma}$ w.r.t. the state $\ket{\psi}$ are $0$ and $1$, where $\hat{n}_{i, \sigma}$ is the occupation number operator acting on spin orbitals. With this we note that $a^\dagger_{a, \sigma} a_{a, \sigma} \ket{\psi_\mathrm{env}} = 1 \ket{\psi_\mathrm{env}}$ since all environment spin orbitals are fully occupied with occupation $1$.
    Further, we will use the following commutation and anticommutation relations:
    \begin{equation} \label{app:eq:comm-anticomm-relations}
        \begin{gathered}
            \{a_{i,\sigma}, a^\dagger_{j,\tau}\}=\delta_{ij} \delta_{\sigma,\tau}\,,\\
            \left[a^{(\dagger)}_{i, \sigma}, a^\dagger_{j, \tau} a_{j, \tau}\right]=0 \quad\mathrm{if}\quad i\neq j \quad\mathrm{or}\quad \tau\neq\sigma\,.
        \end{gathered}
    \end{equation}
    We proceed as follows. We will first split the single- and two-electron sums into multiple sums containing all possible combinations of environment and active orbitals. There we will omit the sums involving an odd number of environment indices since this would change the number of environment electrons which we want to keep constant. Further, we omit the terms where any creation operator acting in the environment space has no annihilation counterpart since this would also change the number of environment electrons. 
    At this point the Hamiltonian is decomposed such that we obtain terms that only act on $\ket{\psi_\mathrm{env}}$ and terms that only act on $\ket{\psi_\mathrm{active}}$. The two-electron matrix elements $h_{tuvw}$ will, additionally, yield mixing terms acting on both $\ket{\psi_\mathrm{env}}$ and $\ket{\psi_\mathrm{active}}$.
    
    We start with the single-electron sum.
    \begin{equation}
        \begin{aligned}
            \sum_{ij,\sigma}h_{ij,\sigma}a^\dagger_{i,\sigma} a_{j,\sigma} \ket{\psi} &= \left(\sum_{tu,\sigma} h_{tu,\sigma}a^\dagger_{t,\sigma} a_{u,\sigma} + \sum_{ab,\sigma}h_{ab,\sigma}a^\dagger_{a,\sigma} a_{b,\sigma}\right)\ket{\psi}\\
            &=\left(\sum_{tu,\sigma} h_{tu,\sigma}a^\dagger_{t,\sigma} a_{u,\sigma} + \sum_{ab,\sigma}h_{ab,\sigma}a^\dagger_{a,\sigma} a_{b,\sigma}\,\delta_{ab}\right)\ket{\psi}\\
            &=\left(\sum_{tu,\sigma} h_{tu,\sigma}a^\dagger_{t,\sigma} a_{u,\sigma} + \sum_{a,\sigma}h_{aa,\sigma}\hat{n}_{a,\sigma}\right)\ket{\psi}\\
            &=\sum_{tu,\sigma} h_{tu,\sigma}a^\dagger_{t,\sigma} a_{u,\sigma}\ket{\psi} + \sum_{a,\sigma}h_{aa,\sigma}\,,
        \end{aligned}
    \end{equation}
    where the Kronecker-delta $\delta_{ab}$ was introduced to explicitly show that if $a\neq b$ the operator $a^\dagger_{a,\sigma} a_{b,\sigma}$ acts as a zero operator since the environment orbitals are fully occupied. When the matrix elements are independent of the spin, we obtain the result used in Equation~\eqref{eq:hamiltonian-many-body-frozen-core}.

    Now, we move on to the two-electron matrix elements. We omit the factor $\frac{1}{2}$ for brevity in the following and add it back at the end of the derivation.
    We start with
    \begin{equation}
        \begin{aligned}
            \sum_{ijkl, \sigma\tau} h_{ijkl, \sigma\tau}\,a^\dagger_{i, \sigma} a^\dagger_{j, \tau} a_{k, \tau} a_{l, \sigma} \ket{\psi} =\bigg[\sum_{\sigma\tau}\bigg(&\sum_{abcd} h_{abcd, \sigma\tau}\,a^\dagger_{a, \sigma} a^\dagger_{b, \tau} a_{c, \tau} a_{d, \sigma}\\
            &+\sum_{tuvw} h_{tuvw, \sigma\tau}\,a^\dagger_{t, \sigma}a^\dagger_{u, \tau}a_{v, \tau}a_{w, \sigma}\\
            &+\sum_{aucw} h_{aucw, \sigma\tau}\,a^\dagger_{a, \sigma}a^\dagger_{u, \tau}a_{c, \tau}a_{w, \sigma}\,\delta_{ac}\delta_{\sigma\tau}\\
            &+\sum_{auvd} h_{auvd, \sigma\tau}\,a^\dagger_{a, \sigma}a^\dagger_{u, \tau}a_{v, \tau}a_{d, \sigma}\,\delta_{ad}\\
            &+\sum_{tbcw} h_{tbcw, \sigma\tau}\,a^\dagger_{t, \sigma}a^\dagger_{b, \tau}a_{c, \tau}a_{w, \sigma}\,\delta_{bc}\\
            &+\sum_{tbvd} h_{tbvd, \sigma\tau}\,a^\dagger_{t, \sigma}a^\dagger_{b, \tau}a_{v, \tau}a_{d, \sigma}\,\delta_{bd}\delta_{\sigma\tau}\bigg)\bigg]\ket{\psi}\,,
        \end{aligned}
    \end{equation}
    where we introduced the Kronecker-deltas to keep the number of environment electrons constant. Now we apply the Kronecker-deltas for all terms and use the relations from Eq.~\eqref{app:eq:comm-anticomm-relations} on the third and sixth terms. Further we use that $\hat{n}_{a,\sigma}$ acts like a scalar $1$ on our state $\ket{\psi_\mathrm{env}\psi_\mathrm{active}}$. We find
    \begin{equation}
        \begin{aligned}
            \sum_{ijkl, \sigma\tau} h_{ijkl, \sigma\tau}\,a^\dagger_{i, \sigma}a^\dagger_{j\tau}a_{k,\tau}a_{l,\sigma}\ket{\psi} =\bigg(& \sum_{abcd, \sigma\tau} h_{abcd, \sigma\tau}\,a^\dagger_{a,\sigma}a^\dagger_{b,\tau}a_{c,\tau}a_{d,\sigma}\\
            &+\sum_{tuvw, \sigma\tau} h_{tuvw, \sigma\tau}\,a^\dagger_{t,\sigma}a^\dagger_{u,\tau}a_{v,\tau}a_{w,\sigma}\\
            &-\sum_{auw,\sigma} h_{auaw, \sigma\sigma}\,a^\dagger_{u, \sigma}a_{w, \sigma}\\
            &+\sum_{auv,\sigma\tau} h_{auva, \sigma\tau}\,a^\dagger_{u, \tau}a_{v, \tau}\\
            &+\sum_{tbw,\sigma\tau} h_{tbbw, \sigma\tau}\,a^\dagger_{t, \sigma}a_{w, \sigma}\\
            &-\sum_{tbv,\sigma} h_{tbvb, \sigma\sigma}\,a^\dagger_{t, \sigma}a_{v, \sigma}\bigg)\ket{\psi}\,.
        \end{aligned}
    \end{equation}
    Applying the swap symmetry (see Eq.~\eqref{app:eq:symm-mat-elems-complex}) to the third and fifth term, and combining same terms yields
    \begin{equation}
        \begin{aligned}
            \sum_{ijkl, \sigma\tau} h_{ijkl, \sigma\tau}\,a^\dagger_{i, \sigma}a^\dagger_{j\tau}a_{k,\tau}a_{l,\sigma}\ket{\psi} =\Bigg\{&\sum_{abcd, \sigma\tau} h_{abcd, \sigma\tau}\,a^\dagger_{a,\sigma}a^\dagger_{b,\tau}a_{c,\tau}a_{d,\sigma}\\
            &+\sum_{tuvw, \sigma\tau} h_{tuvw, \sigma\tau}\,a^\dagger_{t,\sigma}a^\dagger_{u,\tau}a_{v,\tau}a_{w,\sigma}\\
            &+2\sum_{btu, \sigma} \left[ \left(\sum_{\tau}h_{tbbu, \sigma\tau}\right) -  h_{tbub, \sigma\sigma}\right]\,a^\dagger_{t, \sigma}a_{u, \sigma}\Bigg\}\ket{\psi}\,.
        \end{aligned}
    \end{equation}
    Now we need to further investigate the first term. The indices $a,b,c,d$ cannot be all different since we want to keep the number of environment electrons constant. This means that the creation operators need to create the electrons that have been annihilated by the two annihilation operators. This means that
    \begin{equation}
        a^\dagger_{a}a^\dagger_{b}a_{c}a_{d}\ket{\psi}=\left(a^\dagger_{d}a^\dagger_{c}a_{c}a_{d}\delta_{ad}\delta_{bc}+ a^\dagger_{c}a^\dagger_{d}a_{c}a_{d}\delta_{ac}\delta_{bd}\right)\ket{\psi}\,,
    \end{equation}
    and
    \begin{equation}
        \sum_{abcd, \sigma\tau} h_{abcd, \sigma\tau}\,a^\dagger_{a, \sigma}a^\dagger_{b, \tau}a_{c, \tau}a_{d, \sigma}\ket{\psi} =\left( \sum_{ab,\sigma\tau} h_{abba,\sigma\tau}-\sum_{ab, \sigma} h_{abab, \sigma\sigma}\right)\ket{\psi}\,,
    \end{equation}
    where we used the relations from Eq.~\eqref{app:eq:comm-anticomm-relations} and that $\hat{n}_{a,\sigma}$ acts like a scalar $1$ on our state $\ket{\psi}$.
    Combining all terms and adding back the prefactor of $\frac12$ yields
    \begin{equation}
        \begin{aligned}
            \frac{1}{2}\sum_{ijkl, \sigma\tau} h_{ijkl}\,a^\dagger_{i}a^\dagger_{j}a_{k}a_{l}\ket{\psi} =\Bigg\{&+\frac{1}{2}\sum_{tuvw, \sigma\tau} h_{tuvw, \sigma\tau}\,a^\dagger_{t,\sigma}a^\dagger_{u,\tau}a_{v,\tau}a_{w,\sigma}\\
            &+\sum_{btu, \sigma} \left[ \left(\sum_{\tau}h_{tbbu, \sigma\tau}\right) -  h_{tbub, \sigma\sigma}\right]\,a^\dagger_{t, \sigma}a_{u, \sigma}\Bigg\}\ket{\psi}\\
            &+\frac{1}{2}\sum_{ab,\sigma} \left[ \left(\sum_{\tau} h_{abba,\sigma\tau}\right) - h_{abab, \sigma\sigma}\right]\,.
        \end{aligned}
    \end{equation}
    When not considering spin $a^\dagger_{t}a_{u}=\sum_{\sigma}a^\dagger_{t, \sigma}a_{u, \sigma}$ and we obtain the result used in Eq.~\eqref{eq:hamiltonian-many-body-frozen-core}.

\section{Electron configuration in the pseudopotentials} \label{app:electrons-in-pp}
    In this section we list which electrons are valence electrons that are not absorbed in the pseudopotentials used in this work. The electron configuration of an atom is given by the occupation of each one-electron orbital using the principal quantum number ($1,2,3,\ldots$) and the orbital angular momentum ($\mathrm{s},\mathrm{p},\mathrm{d},\ldots$). We follow the notation of the International Union of Pure and Applied Chemistry (IUPAC) \cite{Brett2023Quantities}.
    
    The electrons of the following atoms are valence electrons and are not absorbed in the used norm-conserving pseudopotentials: 
    \begin{itemize}[labelwidth=3.2em, labelsep=1.0em, leftmargin=*, align=left]
        \item[H:] $(1\mathrm{s})^2$
        \item[O:] $(2\mathrm{s})^2$ $(2\mathrm{p})^4$
        \item[Mg:] $(2\mathrm{s})^2$ $(2\mathrm{p})^6$ $(3\mathrm{s})^2$
        \item[S:] $(3\mathrm{s})^2$ $(3\mathrm{p})^4$
        \item[Ti:] $(3\mathrm{s})^2$ $(3\mathrm{p})^6$ $(4\mathrm{s})^2$ $(3\mathrm{d})^2$
        \item[Cr:] $(3\mathrm{s})^2$ $(3\mathrm{p})^6$ $(4\mathrm{s})^2$ $(3\mathrm{d})^4$
        \item[Fe:] $(4\mathrm{s})^2$ $(3\mathrm{d})^6$
        \item[Ni:] $(3\mathrm{s})^2$ $(3\mathrm{p})^6$ $(4\mathrm{s})^2$ $(3\mathrm{d})^8$
        \item[Zn:] $(3\mathrm{s})^2$ $(3\mathrm{p})^6$ $(4\mathrm{s})^2$ $(3\mathrm{d})^{10}$
        \item[Mo:] $(4\mathrm{s})^2$ $(4\mathrm{p})^6$ $(5\mathrm{s})^2$ $(4\mathrm{d})^4$
        \item[Ru:] $(4\mathrm{s})^2$ $(4\mathrm{p})^6$ $(5\mathrm{s})^2$ $(4\mathrm{d})^6$
        \item[Rh:] $(4\mathrm{s})^2$ $(4\mathrm{p})^6$ $(5\mathrm{s})^2$ $(4\mathrm{d})^7$
    \end{itemize}

\section{Active Space scan} \label{app:active_space_scan}

\begin{figure*}[t]
    \centering
    \includegraphics[width=\textwidth]{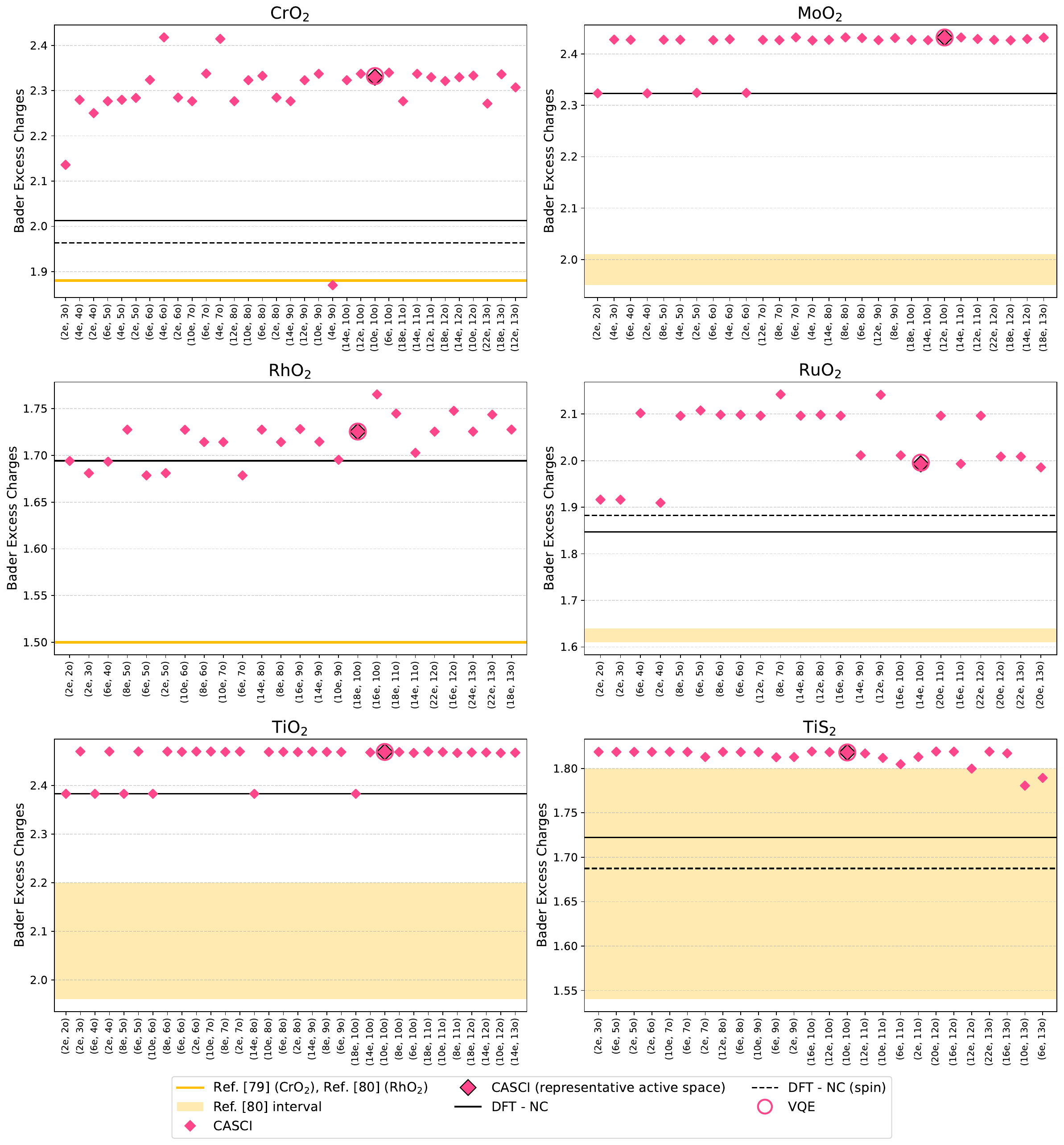}
    \caption{CASCI BECs on the metal atoms of the TMO systems for all scanned active spaces (pink diamonds). The representative active space of each system is marked by a black edge. The VQE results in the representative active spaces are shown as pink circles. The black solid and dashed lines are the spin-unpolarized and spin-polarized DFT-NC BECs, respectively. The yellow lines are the reference values of Ref.~\cite{Lu2018Unraveling} (CrO$_2$) and Ref.~\cite{Choudhuri2020Calculating} (RhO$_2$), and the yellow shaded areas are the reference intervals of Ref.~\cite{Choudhuri2020Calculating}. TiS$_2$ is computed in a cell containing two formula units.}
    \label{fig:as_scan}
\end{figure*}

\begin{figure*}[t]
    \centering
    \includegraphics[width=0.75\textwidth]{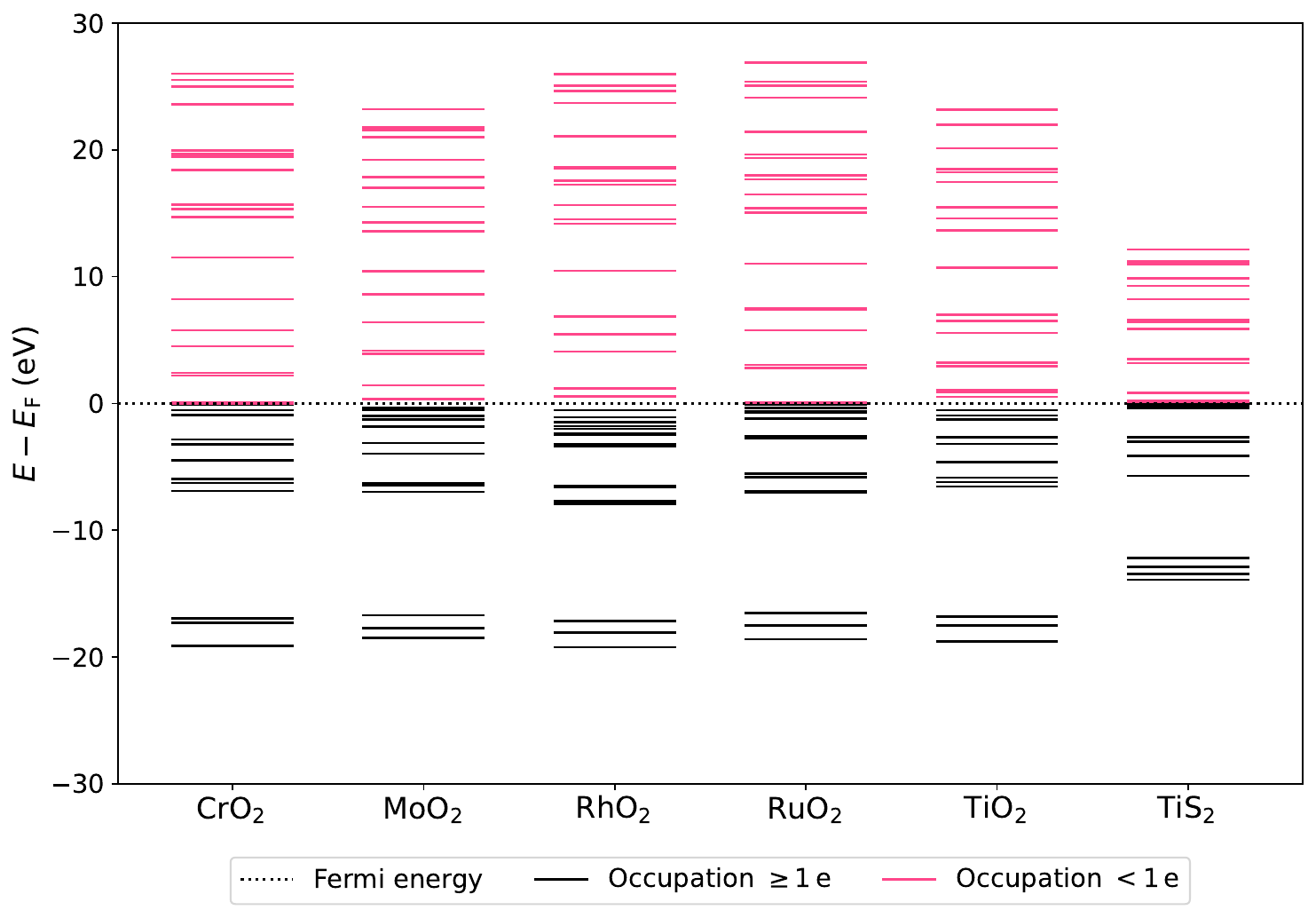}
    \caption{KS energies at the $\brm\Gamma$-point of the spin-unpolarized DFT-NC calculations of the TMO systems relative to the Fermi energy $E_\mathrm{F}$. All computed KS orbitals within $30\,\mathrm{eV}$ of the Fermi energy are shown. Black lines are KS orbitals with an occupation of at least $1\,\mathrm{e}$ and pink lines are KS orbitals with an occupation below $1\,\mathrm{e}$. TiS$_2$ is computed in a cell containing two formula units.}
    \label{fig:ks_energies}
\end{figure*}

In this section we present the CASCI BECs of the TMO systems for all scanned active spaces in Fig.~\ref{fig:as_scan} and the KS energies at the $\brm\Gamma$-point in Fig.~\ref{fig:ks_energies}. In Fig.~\ref{fig:as_scan}, the active spaces of each system are sorted by the number of spatial orbitals and, for equal numbers of spatial orbitals, by the lowest KS orbital in the active space. The representative active spaces used in Fig.~\ref{fig:bader_TMO} and Table~\ref{tab:casci_amplitudes} are marked. For \ce{MoO_2} and \ce{TiO_2}, the BECs fall into two groups. The active spaces whose CASCI ground state consists almost entirely of $\ket{\Psi_0}$ reproduce the DFT-NC BEC, while all other active spaces yield nearly identical BECs. For \ce{CrO_2} and \ce{TiS_2}, the BECs of the active spaces with $11$ to $13$ spatial orbitals vary by $0.07\,\mathrm{e}$ and $0.04\,\mathrm{e}$, respectively. For \ce{RhO_2} and \ce{RuO_2}, the BECs alternate between two groups of values for the largest active spaces and vary by $0.05\,\mathrm{e}$ and $0.11\,\mathrm{e}$, respectively, for the active spaces with $11$ to $13$ spatial orbitals.

\end{document}